\documentclass{ifacconf_amx}

\usepackage{graphicx}      
\usepackage{amsmath,amssymb,amsfonts} 
\usepackage{xcolor}        
\usepackage{cite}          

\usepackage{siunitx}       
\DeclareSIUnit\rpm{rpm}
\DeclareSIUnit\hectare{ha}
\DeclareSIUnit\litre{L}
\DeclareSIUnit\bit{bit}
\usepackage{fancyhdr}      

\usepackage{lipsum}        

\usepackage[scaled]{helvet}
\usepackage[T1]{fontenc}

\newcommand{\ISBN}{UNKNOWN ISBN}

\newcommand{\PaperTitle}{A title for your paper}
\newcommand{\PaperDate}{01 April 1900}
\newcommand{\AuthorFooter}{Oksanen T.}

\makeatletter
\let\old@ssect\@ssect 
\makeatother

\usepackage[hidelinks]{hyperref}      
\usepackage{cleveref}      

\makeatletter
\def\@ssect#1#2#3#4#5#6{%
	\NR@gettitle{#6}
	\old@ssect{#1}{#2}{#3}{#4}{#5}{#6}
}
\makeatother

\usepackage{tikz}
\usepackage{stfloats}
\usepackage{placeins}
\usepackage{todonotes}
\usepackage{booktabs}

\renewcommand{\ISBN}{978-3-911430-14-2}
\renewcommand{\PaperTitle}{EcoTIM: Fuel-saving multi-brand tillage with ISO 11783 TIM}
\renewcommand{\PaperDate}{20 April 2026}
\renewcommand{\AuthorFooter}{Hefele, R. and Oksanen, T.}

\hypersetup{
  pdftitle   = {EcoTIM: Fuel-saving multi-brand tillage with ISO 11783 TIM},
  pdfauthor  = {Ruben Hefele, Karl Th. Renius and Timo Oksanen},
  pdfkeywords= {tractors, ISO 11783, protocol extension, in-vehicle networks, standardization, fuel-efficiency, diesel engine, tractor transmission, multi-brand systems, tillage, ploughs, cultivators, harrows},
  pdfcreator = {LaTeX}
}

\begin{document}
\begin{frontmatter}

   \title{\PaperTitle}

   \author{Ruben Hefele},
   \author{Timo Oksanen}

   \address{Technical University of Munich (TUM), Germany; Professorship of Agrimechatronics, Germany; e-mail: first.last@tum.de; \\
   Munich Institute of Robotics and Machine Intelligence (MIRMI)}

   \begin{abstract}
      Tillage operations account for a large share of on-farm diesel consumption, yet the fuel efficiency of the combined tractor-implement system is not optimised in current practice. Modern continuously variable transmission (CVT) tractors minimise engine fuel consumption internally (``Eco mode''), but they treat the implement as an unknown load and do not account for the effect of vehicle speed on implement draft force. This paper presents EcoTIM, a distributed fuel-optimisation concept in which the tractor and tillage implement cooperate through the extended ISO~11783 (ISOBUS) Tractor Implement Management (TIM) interface to minimise fuel consumption per hectare in real time. In the EcoTIM concept, the tractor Electric Control Unit (ECU) fuses its internal engine, transmission, and traction efficiencies into a single combined efficiency value and its derivative with respect to vehicle speed, and broadcasts both to the implement at the standard \SI{100}{\milli\second} CAN bus cycle. The implement ECU combines these two received scalars with its own analytically known draft force model to evaluate the fuel-consumption gradient, and commands the optimal speed, and as a novel TIM extension, the desired acceleration, back to the tractor. Because only two scalar values are exchanged and neither party discloses proprietary subsystem models, the architecture is inherently multi-brand and plug-and-play. The required data exchange is realised with three new messages and one backward-compatible byte-level extension to the standard TIM speed command, and this paper proposes that these messages are standardised within ISO~11783. The acceleration command enables feed-forward torque and CVT ratio planning on the tractor side, improving transient response compared with speed-only TIM commands. This paper also contains a proof-of-concept simulation with six tillage scenarios and a spatially varying \SI{1}{\kilo\metre} test track for initial concept validation.
   \end{abstract}


\end{frontmatter}

\makeatletter
\newcommand{\bstctlcite}{\@ifnextchar[{\@bstctlcite}{\@bstctlcite[@auxout]}}
\def\@bstctlcite[#1]#2{\@bsphack
  \if@filesw\immediate\write\csname #1\endcsname{\string\citation{#2}}\fi
  \@esphack}
\makeatother
\bstctlcite{IEEEexample:BSTcontrol}

\section{Introduction}
\label{sec:introduction}

Diesel fuel consumption constitutes the single largest variable operating cost in arable farming, and tillage operations, which include ploughing, subsoiling, and secondary cultivation, account for a disproportionately large share of that total because of the high draft forces involved \cite{Renius2020}. A tractor-implement system that is pulling a four-furrow plough at \SI{7}{\kilo\metre\per\hour} in medium-textured soil requires approximately \SIrange{50}{80}{\kilo\watt} of drawbar power, yet only a fraction of the chemical energy in the fuel reaches the soil as useful tillage work. The overall fuel-to-tillage efficiency of a modern \SI{100}{\kilo\watt} tractor is governed by a chain of four distinct sub-efficiencies (engine specific fuel consumption (SFC), transmission mechanical efficiency, wheel-soil tractive efficiency, and the speed-dependent draft force characteristic of the implement), each of which varies non-linearly with the operating point \cite{ASAED497,Renius2020,ZozGrisso2003}.
\par
Modern continuously variable transmission (CVT) tractors already exploit partial knowledge of this efficiency chain. Commercial ``Eco mode'' or powertrain management systems, such as Fendt Tractor Management System, John Deere Efficiency Manager, and CLAAS CMATIC, reduce engine speed along the minimum-SFC locus while the CVT maintains the operator-selected ground speed \cite{Renius2020,Pichlmaier2012}. These systems typically achieve \SIrange{5}{15}{\percent} fuel savings at partial load compared with fixed-throttle operation \cite{Pichlmaier2012}. However, they optimise the tractor powertrain in isolation: the implement is treated as an unknown, variable load, and the relationship between vehicle speed and implement draft force, which for tillage implements such as the moldboard plough is nonlinear \cite{ASAED497}, is not taken into account. Therefore, the resulting operating point is locally optimal for the engine-transmission subsystem but not necessarily for the combined tractor-implement system.
\par
The key information that is missing on the tractor side is the implement's draft force model, and conversely, what the implement lacks is knowledge of the tractor's momentary fuel-to-drawbar efficiency. If these two pieces of information were exchanged in real time, the combined four-component efficiency chain could be optimised on-the-fly, selecting the vehicle speed (and, by extension, the engine-transmission operating point) that minimises fuel consumption per hectare for the prevailing soil conditions and terrain.
\par
The communication infrastructure required for such an exchange already exists. ISO~11783 (ISOBUS) is the established standard for data communication between agricultural tractors and implements, enabling real-time information exchange across machine boundaries. In particular, the Tractor Implement Management (TIM) functionality allows implements to actively command tractor functions such as ground speed, power take-off speed, and hitch position. While this provides a control channel for coordinated operation, current implementations focus primarily on functional control rather than system-wide efficiency optimisation.
\par
However, the current ISO 11783 specification does not define messages for efficiency data exchange. No standard Parameter Group Number (PGN) exists for broadcasting engine SFC, transmission efficiency, tractive efficiency, or any combined efficiency metric from the tractor to the implement. Similarly, no message allows the implement to communicate its draft force model parameters to the tractor. The standard TIM speed command transmits only a speed setpoint, without an acceleration rate that would allow the tractor's powertrain controller to plan smooth torque ramps. These gaps mean that while TIM provides the control channel for speed adjustment, the information channel needed for efficiency-driven optimisation is absent.
\par
Several research projects have addressed tractor-implement fuel optimisation from different angles, yet none has exploited the ISOBUS TIM interface for distributed, multi-brand, real-time cooperative efficiency optimisation. The EKoTech project (BMEL, TU~Braunschweig, 2018 to 2021) demonstrated approximately \SI{10}{\percent} fuel savings through optimised process parameters in ploughing and cultivation, but relied on a centralised controller with full access to all subsystem models, an architecture that is inherently single-brand and not transferable across equipment manufacturers \cite{Troesken2020,EKoTech2017}. Pichlmaier investigated optimal CVT operating strategies and reported \SIrange{3}{8}{\percent} improvement potential, although only the tractor-internal engine-transmission operating point was considered \cite{Pichlmaier2012}. Oksanen and Auernhammer provided a comprehensive review of ISOBUS evolution and identified TIM as the highest automation level in tractor-implement interaction, but did not elaborate on fuel efficiency data exchange \cite{Oksanen2021}. Renius described three automation levels (component control, powertrain management, and TIM), positioning TIM as the natural framework for cooperative tractor-implement control, without however proposing a specific efficiency optimisation protocol \cite{Renius2020}.
\par
This paper presents EcoTIM, a fuel-optimisation concept for multi-brand tractor-implement systems that operates on top of existing ISOBUS TIM infrastructure. The fundamental idea is as follows: the tractor Electric Control Unit (ECU) continuously computes its momentary combined efficiency $\eta_\mathrm{tractor} = \eta_\mathrm{engine} \cdot \eta_\mathrm{transmission} \cdot \eta_\mathrm{tractive}$ together with the derivative $\mathrm{d}\eta/\mathrm{d}v$, and broadcasts both values to the implement via a new message (extension of ISO 11783) proposed in this paper. The implement ECU combines this information with its own draft force model (e.g. based on classic ASAE~D497.7 examples) to evaluate $\mathrm{d}(\text{fuel/ha})/\mathrm{d}v$ and commands the optimal speed, and as a novel TIM extension, the desired acceleration, back to the tractor. The tractor's powertrain controller uses the acceleration command as a feed-forward signal for engine torque and CVT ratio planning, while the speed setpoint serves as a feedback reference. Because only two scalar values ($\eta_\text{tractor}$ and $\mathrm{d}\eta/\mathrm{d}v$) are broadcast and the implement uses only its own draft model for optimisation, neither party needs to disclose proprietary subsystem models, making the architecture inherently multi-brand and plug-and-play.
\par
The contributions of this paper are: (1)~formulation of the EcoTIM distributed optimisation architecture and its ISOBUS message extension; (2)~a proof-of-concept simulation comprising validated submodels for engine SFC, power-split CVT transmission efficiency, Brixius-type tractive efficiency, and ASAE~D497.7 draft force for six tillage scenarios; (3)~a \SI{1}{\kilo\metre} virtual test track with spatially varying soil conditions and terrain; (4)~a simulated ISOBUS message exchange at \SI{100}{\milli\second} bus rate that demonstrates real-time speed adaptation; and (5)~quantification of fuel savings relative to the realistic baseline of a modern CVT tractor operating in Eco mode at constant operator-selected speed.

\section{State of the Art}
\label{sec:sota}

The fuel consumption of a tractor and tillage implement system is determined by a chain of four sub-efficiencies: diesel engine fuel conversion, mechanical transmission, wheel-to-soil traction, and the speed-dependent draft force of the implement. Each of these subsystems has been studied extensively in isolation, yet their combined, real-time optimisation through standardised data exchange remains an open problem. This section reviews the state of the art for each subsystem and for the ISOBUS communication infrastructure that could enable cooperative optimisation.

\subsection{ISOBUS and Tractor Implement Management}
\label{sec:bg_isobus}

ISOBUS is the established serial data network for agricultural tractors and implements, developed from the German LBS standard first proposed in 1987 and subsequently adopted as an international standard in the early 2000s \cite{Oksanen2021,ISO11783-1}. The protocol is based on SAE J1939 \cite{J1939} over a CAN~2.0B physical layer at \SI{250}{\kilo\bit\per\second}, with cyclic message broadcast at \SI{100}{\milli\second} intervals for real-time process data \cite{ISO11783-7}. ISOBUS defines a comprehensive set of PGNs for engine state, wheel speed, ground speed, and rear hitch state including draft force \cite{J1939-71,ISO11783-7}.
\par
TIM, defined in ISO~11783-9 \cite{ISO11783-9} and related guidelines, allows the implement ECU to command tractor functions in a closed-loop manner. TIM was historically referred to as ``Class~3'' functionality, indicating that the implement not only receives data from the tractor (Class~1) and is controlled by the tractor (Class~2), but actively commands tractor functions \cite{Renius2020,Oksanen2021}. The current TIM specification supports speed setpoints, PTO speed, rear and front hitch position, and auxiliary hydraulic valve commands. 
\par
However, neither ISOBUS nor TIM in its current form defines messages for efficiency data exchange. No standard PGN exists for broadcasting engine SFC, transmission efficiency, tractive efficiency, or any combined efficiency metric from the tractor to the implement. The standard TIM speed command transmits only a speed setpoint, without an acceleration rate that would enable the tractor's powertrain controller to plan smooth torque ramps. Similarly, no message allows the implement to communicate its draft force model parameters to the tractor.
\subsection{Diesel engine fuel consumption models}
\label{sec:bg_engine}

The specific fuel consumption (SFC, denoted $b_e$ following Renius \cite{Renius2020}) of a diesel engine depends on both the engine load and its operating speed. ASAE~D497.7 \cite{ASAED497} provides a widely used empirical model in which the volumetric SFC is expressed as a function of the load fraction
\begin{equation}
	\chi = \frac{P}{P_\mathrm{rated}},
\end{equation}
where $P$ is the instantaneous engine power and $P_\mathrm{rated}$ is the rated engine power. The model also includes a partial-throttle multiplier $f_\mathrm{pt}$, which accounts for the efficiency gains associated with operating the engine at reduced speed while delivering the same power:

\begin{equation}
	b_e = b_{e,\mathrm{base}} \cdot \left(0.22 + \frac{0.096}{\chi}\right) \cdot f_\mathrm{pt},
	\label{eq:asae_sfc}
\end{equation}

\noindent where $b_{e,\mathrm{base}}$ is a scaling factor that calibrates the fleet-average model to a specific engine. The partial-throttle factor $f_\mathrm{pt}$ captures the speed-dependent efficiency of the engine itself, independent of the transmission.
\par
For modern Stage~IV/V engines with common-rail injection and variable-geometry turbocharging, specific fuel consumption varies with engine speed and load. For example, the Fendt~514~Vario~S4, used as a reference tractor in this study, exhibited \SI{223}{\gram\per\kilo\watt\per\hour} at peak torque ($\approx$~\SI{1400}{\rpm}) and \SI{259}{\gram\per\kilo\watt\per\hour} at rated speed ($\approx$~\SI{2000}{\rpm}) as measured at the PTO shaft \cite{DLG2016PowerMix}, reflecting a \SI{36}{\gram\per\kilo\watt\per\hour} increase due to the engine's speed dependence. Under lower partial-load conditions (\SI{90}{\percent} of rated speed and \SI{40}{\percent} of rated power), SFC rises further to about \SI{298}{\gram\per\kilo\watt\per\hour}, illustrating the broader operating range of the engine under realistic field conditions.
\par
An alternative to the ASAE hyperbolic model is the Willans line representation:
\begin{equation}
	\dot{m}_f = \dot{m}_{f,0}(n) + k_f(n)\, P,
\end{equation}
where $\dot{m}_f$ denotes the fuel mass flow rate, $n$ the engine speed, $\dot{m}_{f,0}(n)$ the speed-dependent zero-load (friction) fuel consumption, and $k_f(n)$ the incremental fuel consumption per unit power \cite{Guzzella2010}. This formulation inherently captures the speed dependence of the SFC at full load, which is not represented in the ASAE model, since the base curve in~(\cref{eq:asae_sfc}) depends solely on the load fraction~$\chi$.

\subsection{Transmission efficiency}
\label{sec:bg_transmission}

Modern European tractors above \SI{100}{\kilo\watt} predominantly use power-split CVTs, which combine a hydrostatic variator with a mechanical planetary path \cite{Renius2020}. Kress \cite{Kress1968} classified power-split structures into output-coupled (OC), input-coupled (IC), and compound-coupled (CC) categories. The Fendt Vario, the first mass-produced tractor power-split CVT (series production from 1996), uses an OC structure with two synchronised ranges and 45-degree bent-axis hydrostatic units \cite{Renius2020}. The ZF Terramatic (originally ``Eccom'', introduced with Deutz-Fahr tractors in 2001, later also CLAAS from 2013) employs an IC structure with four synchronised ranges \cite{Renius2020}.

Transmission efficiency depends on the hydrostatic power fraction, which varies with the ratio of vehicle speed to the synchronous (lock-up) speed of each range. At the lock-up point, all power flows through the mechanical path and variator losses are zero; away from lock-up, an increasing fraction passes through the hydrostatic units with typical variator efficiencies of $0.85$ to $0.96$ depending on the unit type and operating pressure \cite{Renius2020,Reick2018}. For the OC structure, Renius \cite{Renius2020} demonstrated that total efficiency remains ``very flat'' across a wide speed range if the variator efficiency exceeds $0.85$ (fig.~5.64 in \cite{Renius2020}). Measured full-load transaxle efficiency of the Fendt Vario ML~200 was reported as \SI{82}{\percent} at \SI{4}{\kilo\metre\per\hour} rising to \SI{84}{\percent} at \SIrange{6}{12}{\kilo\metre\per\hour} (fig.~5.69 in \cite{Renius2020}), which is consistent with the lock-up point being located near the primary tillage speed range.

For the IC structure, Ziegler et~al. \cite{Ziegler2014} calculated power losses for the first range of the ZF Terramatic~11 and reported lowest losses at \SI{10}{\kilo\metre\per\hour}, where hydrostatic power is zero. This confirms that both major CVT architectures place a lock-up point in or near the \SIrange{8}{12}{\kilo\metre\per\hour} tillage working range.

The only complete published efficiency map (speed versus torque) for a tractor transmission remains the Reiter \cite{Reiter1990} measurement of a Fendt~310~LSA stepped transmission (reproduced in Renius \cite{Renius2020}, fig.~5.138), showing \SI{88}{\percent} at full load decreasing to \SI{84}{\percent} at \SI{50}{\percent} load and \SI{75}{\percent} at \SI{15}{\percent} load. Partial-load CVT data is not publicly available, which represents a gap for accurate simulation; however, component-level efficiencies from Reick \cite{Reick2018} (spur gear $0.99$, planetary $0.98$, variator $0.86$, rear axle $0.96$) enable analytical modelling that has been validated against DLG PowerMix fuel consumption data \cite{DLG2016PowerMix} within \SI{3.2}{\percent} error.

\subsection{Traction models}
\label{sec:bg_traction}
The conversion of mechanical torque at the wheel hub into drawbar pull is governed by the wheel--soil interaction, characterised by wheel slip, motion resistance, and the resulting tractive efficiency. The Brixius model \cite{Brixius1987}, adopted in ASAE~D497.7 \cite{ASAED497}, expresses the net traction coefficient and motion resistance as functions of a dimensionless wheel mobility number~$B_n$ that combines cone index~(CI), tyre dimensions, and wheel load. Zoz and Grisso \cite{ZozGrisso2003} recalibrated the Brixius coefficients for modern radial tyres, giving
\begin{equation}
	\kappa = 0.88\,\bigl(1 - e^{-0.08\,B_n}\bigr)\,\bigl(1 - e^{-7.0\,s}\bigr) - \rho
\end{equation}
\noindent where $s$ is the wheel slip ratio and $\rho$ the motion resistance coefficient. For radial tyres,
\begin{equation}
	\rho = \frac{1.2}{B_n} + \frac{0.5\,s}{\sqrt{B_n}} + 0.03\,.
\end{equation}
\noindent Tractive efficiency follows as
\begin{equation}
	\eta_{\mathrm{tractive}} = \frac{\kappa}{\kappa + \rho}\,(1 - s)\,.
\end{equation}
\noindent For a four-wheel-drive~(4WD) tractor operating on tilled agricultural soil (CI~$\approx$~\SI{900}{\kilo\pascal}), optimal tractive efficiency of \SIrange{65}{70}{\percent} is typically attained at \SIrange{8}{12}{\percent} slip \cite{ZozGrisso2003,Renius2020}.
\par
The traction model introduces a soil-condition dependency that is not known 
\textit{a priori} and varies across the field. Cone index depends on soil type, 
moisture content, and cultivation history, with typical values ranging from 
\SI{450}{\kilo\newton\per\metre\squared} for wet soft or loose sandy soils to
\SI{1800}{\kilo\newton\per\metre\squared} for heavy dry uncultivated soil \cite{Renius2020}. 
This spatial variability is one of the motivations for real-time, on-the-fly 
optimisation rather than pre-computed static setpoints.


\subsection{Implement draft force models}
\label{sec:bg_pullingforce}

ASAE~D497.7~\cite{ASAED497} provides a widely used empirical draft force model
for tillage implements:
\begin{equation}
	F_{\mathrm{draft}} = f_{\mathrm{s}} \!\cdot\!\bigl[A + B\,v + C\,v^{2}\bigr]\,w\,d
	\label{eq:draft}
\end{equation}
where $f_{\mathrm{s}}$ is a dimensionless soil-texture adjustment
factor ($1.0$ for fine/clay, $0.70$ for medium, $0.45$ for coarse),
$A$, $B$, $C$ are implement-specific coefficients tabulated for more
than 30~implement types (in the ASAE unit system: $A$ in \si{\newton\per\metre\per\centi\metre}, $B$ in \si{\newton\hour\per\metre\per\kilo\metre\per\centi\metre}, $C$ in \si{\newton\hour\squared\per\metre\per\kilo\metre\squared\per\centi\metre}),
$v$ is the forward speed in~\si{\kilo\metre\per\hour},
$w$ is the working width in~\si{\metre},
and $d$ is the tillage depth in~\si{\centi\metre}. All $A$, $B$, $C$ values cited hereafter use these units.
For a mouldboard plough the quadratic speed term dominates at
field speeds ($C=5.1$ in the consistent ASAE unit system),
causing $F_{\mathrm{draft}}$ to rise sharply with velocity.
For disc harrows and tine cultivators the linear term~($B$) is
the dominant contributor.
\par
The EKoTech project (BMEL, TU~Braunschweig) validated the
ASAE~D497.7 model under European tillage conditions and
demonstrated that optimised process parameters (speed, depth
and engine setpoint) can reduce fuel consumption by
approximately \SI{10}{\percent} during ploughing and
cultivation~\cite{Troesken2020,EKoTech2017}.
Ahokas and Oksanen~\cite{Ahokas2015} provided per-tine draft
force data for European spring-tine harrows that complement
the ASAE tabulation, which was developed primarily for
North~American implements and soils.
Reported values range from
\SI{80}{\newton\per tine}
($d=\SI{5}{\centi\metre}$, $v=\SI{8}{\kilo\metre\per\hour}$,
humus soil) to
\SI{182}{\newton\per tine} (clay soil) for S-tine implements at
typical secondary-tillage depths~\cite{Ahokas2015}.
\par
The key insight for EcoTIM is that the implement controller
stores its own $A$, $B$, $C$, $f_{\mathrm{s}}$ coefficients and current working parameters ($w$, $d$), and can therefore estimate the draft force gradient $\mathrm{d}F_{\mathrm{draft}}/\mathrm{d}v$ analytically.
When combined with the tractor's broadcast efficiency gradient $\mathrm{d}\eta/\mathrm{d}v$, the implement can determine the direction and magnitude of the speed change that minimises fuel per hectare, without requiring access to any tractor-internal model. R\"{o}\ss{}ler et~al.~\cite{Roessler2012} have shown that
these model parameters can be identified online during field
operation using a cluster-based approach, even when soil
conditions vary within a field. Naturally, the implement engineers may also use machine
learning principles, including neural networks, to model the
draft force estimate and its gradient.

\section{Proposed Solution}
\label{sec:solution}

This section describes the complete EcoTIM concept as an engineering specification on both the tractor and implement side.

\subsection{System overview}
\label{subsec:overview}

EcoTIM is a distributed, multi-brand fuel optimisation architecture in which the tractor and implement cooperate through a minimal data exchange to minimise fuel consumption per hectare in real time. The architecture partitions the optimisation problem between two independent subsystems:

\begin{itemize}
\item The \textbf{tractor ECU} knows its own engine map, transmission state, and traction conditions. It fuses these three sub-efficiencies into a single combined efficiency $\eta_\mathrm{tractor} = \eta_\mathrm{engine} \cdot \eta_\mathrm{transmission} \cdot \eta_\mathrm{tractive}$ and computes the derivative $\mathrm{d}\eta/\mathrm{d}v$. It broadcasts both values every \SI{100}{\milli\second}.
\item The \textbf{implement ECU} knows its own ASAE~D497.7 draft model coefficients ($A$, $B$, $C$, $f_s$, $w$, $d$), or any other similar proprietary model. It receives $\eta$ and $\mathrm{d}\eta/\mathrm{d}v$ from the tractor, combines them with its own $F_\text{draft}(v)$ and $\mathrm{d}F_\text{draft}/\mathrm{d}v$, and commands the optimal speed and acceleration back to the tractor via TIM.
\end{itemize}

Neither party needs the other's internal model. The tractor does not know the implement type, and the implement does not know the engine map or transmission type. This separation of concerns is what enables multi-brand interoperability and distinguishes EcoTIM from centralised approaches such as EKoTech \cite{Troesken2020}.

\begin{figure}[h!]
	\centering
	\includegraphics[width=\columnwidth]{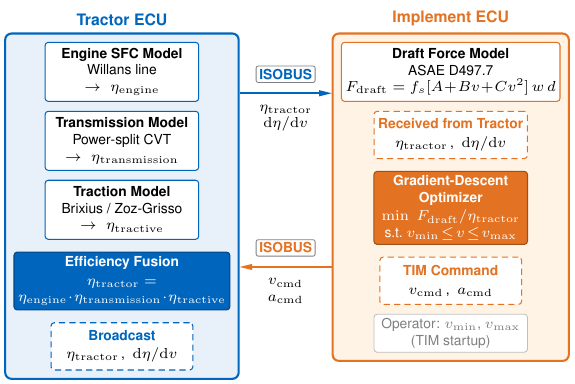}
	\caption{EcoTIM system architecture. The tractor ECU fuses three internal submodels into a combined efficiency and broadcasts it via ISOBUS. The implement ECU combines this with its own draft force model to command the optimal speed and acceleration.}
	\label{fig:system_overview}
\end{figure}

The overall fuel consumption per hectare can be expressed as proportional to $F_\text{draft}(v) / \eta_\mathrm{tractor}(v)$, since area rate is proportional to $v \cdot w_\mathrm{eff}$ and drawbar power is $F_\text{draft} \cdot v$. EcoTIM seeks the vehicle speed $v$ that minimises this ratio within operator-defined limits $[v_\mathrm{min}, v_\mathrm{max}]$. \Cref{fig:system_overview} illustrates the architecture and data flow between the two ECUs.

\subsection{Tractor side: Eco mode and efficiency computation}
\label{subsec:tractor_side}

Modern CVT tractors already implement an ``Eco mode'' in which the transmission controller selects the engine speed that minimises SFC for the current power demand \cite{Renius2020,Pichlmaier2012}. In this mode, the CVT decouples engine speed from ground speed, allowing the controller to sweep from peak-torque speed $n_{T,\mathrm{max}}$ to rated speed $n_\mathrm{rated}$ and select the speed at which the required torque is available with minimum SFC. Eco mode uses only the engine fuel map; it does not consider transmission efficiency or traction conditions.

EcoTIM does not replace Eco mode. Instead, it operates on top of Eco mode by adding vehicle speed optimisation. The tractor ECU continues to manage the engine-transmission operating point via Eco mode (or any other internal powertrain management strategy), while additionally computing and broadcasting its total efficiency state. 

\newpage The tractor-side computation at each \SI{100}{\milli\second} cycle is:
\begin{enumerate}
\item Compute the current combined efficiency: $\eta_\mathrm{tractor} = (b_{e,\mathrm{best}} / b_{e,\mathrm{actual}}) \cdot \eta_\mathrm{transmission} \cdot \eta_\mathrm{tractive}$, where $b_{e,\mathrm{best}}$ is the minimum achievable SFC and $\eta_\mathrm{tractive}$ is the tractive efficiency.
\item Estimate the derivative $\mathrm{d}\eta/\mathrm{d}v$ by evaluating the tractor's internal models at two perturbed speeds and applying a central-difference approximation. With a perturbation step $\delta v = \SI{0.3}{\kilo\metre\per\hour}$, this gives
\begin{equation}
	\frac{\mathrm{d}\eta}{\mathrm{d}v} \approx \frac{\eta_\text{tractor}(v + \delta v) - \eta_\text{tractor}(v - \delta v)}{2\,\delta v}.
\end{equation}
\item Broadcast $\eta_\mathrm{tractor}$ and $\mathrm{d}\eta/\mathrm{d}v$ via the EcoTIM Proprietary~A message.
\end{enumerate}

The perturbation evaluation uses the tractor's own engine, transmission, and traction models, which should already be available in the tractor ECU for powertrain management embedded software. No additional sensors are required; the tractor simply computes what its efficiency would be if speed changed from the current operating point.

\subsection{Implement side: optimizer algorithm}
\label{subsec:optimizer}

The implement-side optimizer determines the fuel-optimal vehicle speed
using only its own ASAE draft model and two scalar values received from
the tractor ($\eta_{\mathrm{tractor}}$ and
$\mathrm{d}\eta/\mathrm{d}v$, where $\mathrm{d}\eta/\mathrm{d}v$
is understood to refer to $\eta_{\mathrm{tractor}}$ throughout).
The algorithm, executed every \SI{100}{\milli\second}, proceeds as follows:

\textbf{Step~1: Draft evaluation.}
The implement computes draft force at three speeds using its own
analytical model (cf.~\cref{eq:draft}):
\begin{align}
	F_{\mathrm{draft}}(v) &= f_{\mathrm{s}} \cdot \bigl[A + B\,v + C\,v^{2}\bigr]\, w\, d \\
	F_{\mathrm{draft}}(v \pm \delta v) &= f_{\mathrm{s}} \cdot \bigl[A + B\,(v \pm \delta v)
	+ C\,(v \pm \delta v)^{2}\bigr]\, w\, d
\end{align}
where $\delta v$ is a small probe step, e.g.\ \SI{0.5}{\kilo\metre\per\hour}.

\textbf{Step~2: Efficiency linearisation.}
The implement estimates tractor efficiency at the probe speeds using
the broadcast derivative:
\begin{equation}
	\eta_{\mathrm{tractor}}(v \pm \delta v) \approx \eta_{\mathrm{tractor,rx}}
	\pm \frac{\mathrm{d}\eta}{\mathrm{d}v}\bigg|_{\mathrm{rx}}
	\cdot \delta v
\end{equation}

\textbf{Step~3: Gradient computation.}
The fuel proxy $F_{\mathrm{draft}}/\eta_{\mathrm{tractor}}$ is evaluated at three
points and the central-difference gradient is computed:
\begin{equation}
	\raisebox{1.5em}{$\displaystyle
		\frac{\mathrm{d}}{\mathrm{d}v}\!\left(\frac{F_{\mathrm{draft}}}{\eta_{\mathrm{tractor}}}\right)
		\approx
		\frac{1}{2\,\delta v}
		\!\left[
		\frac{F_{\mathrm{draft}}(v+\delta v)}{\eta_{\mathrm{tractor}}(v+\delta v)}
		-
		\frac{F_{\mathrm{draft}}(v-\delta v)}{\eta_{\mathrm{tractor}}(v-\delta v)}
		\right]
		$}
	\label{eq:gradient}
\end{equation}

\textbf{Step~4: Speed command.}
The normalised relative gradient determines the speed step:
\begin{equation}
	\Delta v = -k_{v} \cdot
	\frac{\mathrm{d}(F_{\mathrm{draft}}/\eta_{\mathrm{tractor}})/\mathrm{d}v}
	{F_{\mathrm{draft}}/\eta_{\mathrm{tractor}}},
	\qquad
	|\Delta v| \leq \Delta v_{\mathrm{max}}
\end{equation}
where $k_{v}$ is a tunable velocity gain (e.g.\
\SI{2.0}{\kilo\metre\per\hour} per unit relative gradient) and $\Delta v_{\mathrm{max}}$ is
a safety limit on the maximum step size per cycle (e.g.\ \SI{0.5}{\kilo\metre\per\hour}). The new
speed command is
$v_{\mathrm{cmd}} = \mathrm{clip}(v_{\mathrm{actual}} + \Delta v,\;
v_{\mathrm{min}},\; v_{\mathrm{max}})$.

\begin{minipage}{\columnwidth}
	\textbf{Step~5: Acceleration command.}
	The acceleration setpoint is derived from the same normalised gradient used in Step~4: 
	\begin{equation}
		a_{\mathrm{cmd}} = -k_{a} \cdot
		\frac{\mathrm{d}(F_{\mathrm{draft}}/\eta_{\mathrm{tractor}})/\mathrm{d}v}
		{F_{\mathrm{draft}}/\eta_{\mathrm{tractor}}},
		\qquad
		|a_{\mathrm{cmd}}| \leq a_{\mathrm{max}}
	\end{equation}
	\vspace{0.001 cm}
\end{minipage}
where $k_{a}$ is a tunable acceleration gain and $a_{\mathrm{max}}$ is a safety limit (e.g.\ \SI{0.5}{\metre\per\second\squared}).
A large gradient (far from optimum or entering a new soil zone)
produces a large $|a_{\mathrm{cmd}}|$, signalling the tractor to respond urgently. A small gradient (near optimum) produces a gentle acceleration, signalling fine-tuning.

The entire procedure requires three draft-force evaluations
(analytical, no iteration) and one linear interpolation. This is
computationally trivial and fits within a single \SI{100}{\milli\second} bus tick on any
standard implement ECU.

The optimizer algorithm in the implemented ECU may be considerably more sophisticated, without requiring modifications to the EcoTIM concept or its interfaces. The five-step approach presented here serves as a baseline reference algorithm and is the one applied in the remainder of this paper.

\subsection{Tractor speed and acceleration response}
\label{subsec:tractor_response}

The tractor's powertrain controller uses the speed command $v_\mathrm{cmd}$ and acceleration command $a_\mathrm{cmd}$ from the implement to adjust vehicle speed. The response model combines feed-forward (from $a_\mathrm{cmd}$) and feedback (from speed error):

\begin{equation}
v_\mathrm{new} = v_\mathrm{actual} + \underbrace{a_\mathrm{cmd} \cdot \Delta t}_\text{feed-forward} + \underbrace{\alpha \cdot (v_\mathrm{cmd} - v_\mathrm{actual})}_\text{feedback}
\label{eq:dynamics}
\end{equation}

\noindent where $\alpha = \Delta t / \tau$ is the feedback gain and $\tau$ is the tractor's speed response time constant. A system identification on a Fendt~314~Vario CVT tractor \cite{Soitinaho2023} yielded $\tau_v = \SI{3.72}{\second}$ under no-load conditions; under primary-tillage draft the effective response is faster because the draft force damps deceleration and the powertrain controller operates at higher torque-output gain. Therefore, the estimated value $\tau = \SI{2}{\second}$ is adopted for the simulation. When $|a_\mathrm{cmd}|$ is large, the feed-forward term dominates and the tractor responds quickly to zone changes. When $|a_\mathrm{cmd}|$ is near zero, the feedback term provides gentle correction toward the setpoint.

The feed-forward term enables the tractor's engine controller to ramp 
torque proactively (reducing turbo lag) and the CVT controller to 
pre-compute the ratio trajectory (reducing hydraulic pressure transients 
in the hydrostatic unit). Without $a_\mathrm{cmd}$, the tractor can only 
react to speed error after it develops, which is inherently slower and 
leads to overshoot.

If the engine cannot physically sustain the commanded speed at the current soil conditions (engine overloaded or excessive wheel slip), the tractor reduces speed to the maximum feasible level. This power-limiting behaviour is transparent to the implement; the tractor simply does not reach the commanded speed, and the implement observes the lower actual speed and reduced $\eta_\text{tractor}$ in the next broadcast cycle, adjusting its command accordingly.

The operator retains ultimate authority through minimum and maximum speed limits communicated at session startup, and through direct override immediately suspends TIM commands per ISO~11783-9 \cite{ISO11783-9} and related guidelines.

\subsection{ISOBUS message definitions}
\label{subsec:messages}

EcoTIM requires one new message, which can be deployed with minor modification to the ISO~11783 standard, and one backward-compatible extension to the standard TIM speed command. \Cref{tab:messages} summarises the complete message architecture.

\begin{table*}[b]
\centering
\caption{EcoTIM message architecture. Standard ISO~11783 messages (top) and proposed EcoTIM extensions (bottom).}
\label{tab:messages}
\small
\begin{tabular}{lllll}
\hline
PGN & Name & Direction & Rate & Content \\
\hline
65096 & Ground-Based Speed & Tractor $\rightarrow$ All & 100~ms & Machine speed, distance \\
65094 & Rear Hitch State & Tractor $\rightarrow$ All & 100~ms & Hitch position, draft force \\
65098 & TIM Speed Command & Impl $\rightarrow$ Tractor & 100~ms & Speed setpoint \\
\hline
New PGN & EcoTIM Efficiency Broadcast & Tractor $\rightarrow$ Impl & 100~ms & $\eta_\mathrm{tractor}$, $\mathrm{d}\eta/\mathrm{d}v$ \\
TIM command ext. & TIM Speed + Acceleration (extended) & Impl $\rightarrow$ Tractor & 100~ms & Speed + acceleration \\
\hline
\end{tabular}
\end{table*}

The \textbf{Tractor Efficiency Broadcast} carries the fused efficiency $\eta_\mathrm{tractor}$ as a percentage and the derivative $\mathrm{d}\eta/\mathrm{d}v$ in \si{\percent\second\per\metre}. The individual sub-efficiencies and engine load fraction are included for diagnostics but are not required by the optimizer.

The \textbf{TIM Speed and Acceleration Command} standard TIM speed command is extended by placing the acceleration setpoint in reserved bytes of the 8-byte CAN frame (resolution \SI{0.001}{\metre\per\second\squared}). This extension is backward-compatible: a non-EcoTIM tractor ignores the reserved bytes, while an EcoTIM-capable tractor uses them for feed-forward control.

The total additional bus load from the two EcoTIM messages is approximately 10--20 frames per second, corresponding to less than \SI{1}{\percent} of the \SI{250}{\kilo\bit\per\second} CAN capacity~\cite{Oksanen2021}, which is well within the available bandwidth on a typical ISOBUS installation.

\section{Simulation Study}
\label{sec:results}

A proof-of-concept simulation was developed to demonstrate the EcoTIM concept described in \cref{sec:solution}. The simulation comprises four textbook-based and inference calibrated submodels for the tractor-implement efficiency chain, an ISOBUS message exchange mockup that implements the protocol of \cref{tab:messages}, a \SI{1}{\kilo\metre} virtual test track with spatially varying conditions, and six tillage scenarios covering a representative range of European draft-only implements.

\subsection{Simulation setup}
\label{subsec:sim_setup}

The simulation models a \SI{100}{\kilo\watt} class 4WD tractor based on the Fendt~514 Vario~S4 (DLG PowerMix Test 2016) with an OC power-split CVT. The tractor is parameterised from published DLG dynamometer data (10 operating points including idle) and from component-level efficiencies reported by Reick \cite{Reick2018}. The submodels are chained in the power flow direction: implement draft force $\rightarrow$ traction $\rightarrow$ transmission $\rightarrow$ engine $\rightarrow$ fuel consumption. An iterative convergence loop (typically 3 to 5 iterations) resolves the coupling between transmission efficiency and engine operating point.

The ISOBUS message exchange is implemented as a time-stepping simulation at \SI{100}{\milli\second} cycle time, in which tractor and implement functions exchange data through struct-based message objects that carry real PGN numbers and Suspect Parameter Number (SPN) fields per ISO~11783-7 \cite{ISO11783-7}. CAN frame encoding, addressing, priority arbitration, and transport protocol are not modelled, since they do not affect the optimisation logic of this simulation setup. 

The baseline for comparison is a modern CVT tractor operating in Eco mode at constant wheel-based speed. In this mode, the tractor ECU sweeps engine speed from $n_{T,\mathrm{max}}$ to $n_\mathrm{rated}$, checks torque availability on the full-load curve, and selects the engine speed that minimises SFC for the current power demand \cite{Renius2020}. This is the realistic operating condition of modern tractors such as the Fendt Vario (TMS), John Deere (Efficiency Manager), and CLAAS (CMATIC); it is not a naive fixed-throttle assumption. The simulation is not exactly any of the commercial implementations, but serves sufficiently for the baseline comparison.

For each scenario, the \SI{1}{\kilo\metre} track is traversed at nine constant speed setpoints (\SIrange{4}{12}{\kilo\metre\per\hour} in \SI{1}{\kilo\metre\per\hour} steps) with Eco mode active, and once with EcoTIM enabled. The nine baseline runs map the full fuel-versus-time trade-off curve for each implement (\cref{fig:fuel_vs_time}). From this sweep, a scenario-specific operator speed is selected as the primary comparison point for \cref{tab:fuel_comparison} (for example, \SI{8}{\kilo\metre\per\hour} for a well-matched plough, \SI{12}{\kilo\metre\per\hour} for a light disc cultivator), representing the speed a skilled operator would realistically choose.

If the tractor cannot maintain a given speed setpoint at a particular track position (engine overloaded or slip exceeding \SI{25}{\percent}), the speed is automatically reduced to the maximum feasible level using a bisection search. Total fuel consumption (litres), time per hectare (\si{\minute\per\hectare}), and fuel per hectare (\si{\litre\per\hectare}) are computed by integrating along the track for each run. Tillage quality is out of scope; all speeds within the operator-defined range are considered acceptable. 

\subsection{Submodels}
\label{subsec:submodels}

\textbf{Engine fuel consumption.}
The engine fuel map is based on a two-dimensional Willans line model calibrated to ten DLG PowerMix operating points of the Fendt~514 Vario~S4 \cite[p.~3]{DLG2016PowerMix}, including idle. Since DLG measures shaft power at the PTO stub, all power values were first rescaled to crankshaft level by dividing by a PTO drivetrain efficiency $\eta_\mathrm{PTO} = 0.93$, while the measured fuel flow rates remain unchanged. This factor follows from the Deutz TCD~4.1~L4~Agri data sheet \cite{Deutz_TCD41L4}, which states a crankshaft minimum of $b_e = \SI{208}{\gram\per\kilo\watt\per\hour}$; the DLG PTO minimum is \SI{223}{\gram\per\kilo\watt\per\hour}, giving $\eta_\mathrm{PTO} = 208/223 = 0.93$. As a cross-check, Renius \cite{Renius2020} (fig.~4.23) reports $b_e \approx$~\SI{205}{\gram\per\kilo\watt\per\hour} for the related Deutz TCD~2013~L06~4V (6-cylinder, \SI{7.1}{\litre}, Stage~IV), which would imply $\eta_\mathrm{PTO} = 0.92$, consistent with a slightly more efficient larger engine. The full-load power envelope was rescaled with the same factor.

\begin{figure}[h]
	\centering
	\includegraphics[width=\columnwidth]{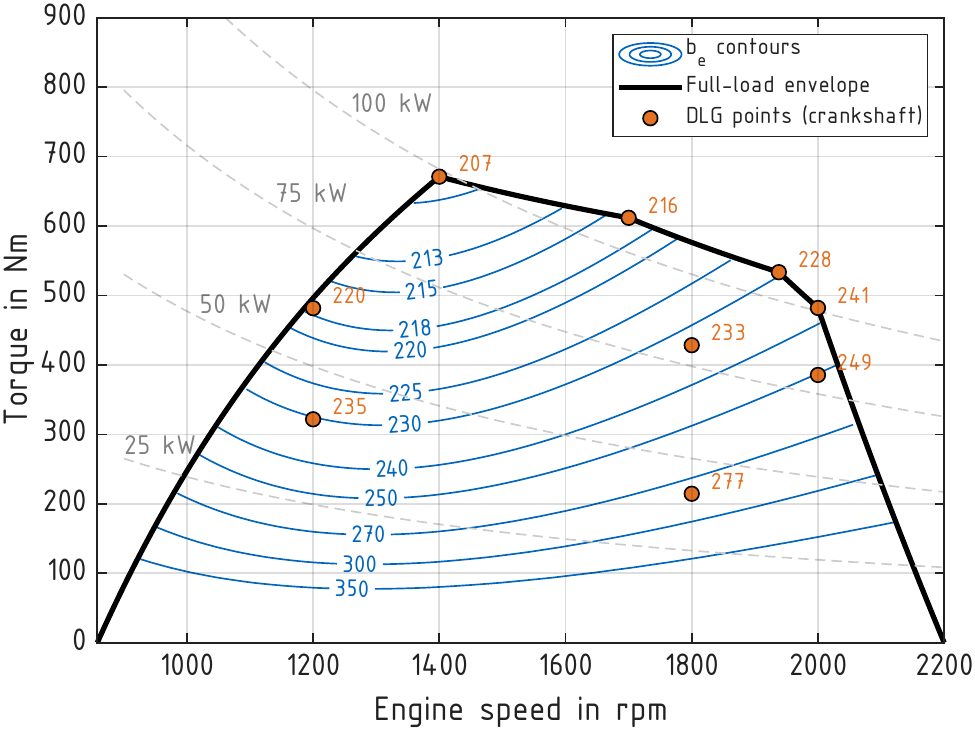}
	\caption{Engine $b_e$ map (Willans line, crankshaft-referenced after PTO correction with $\eta_\mathrm{PTO} = 0.93$). Contour lines show constant $b_e$ in \si{\gram\per\kilo\watt\per\hour}. Orange dots: DLG measured operating points rescaled to crankshaft power, labelled with the corrected $b_e$. Thick line: full-load torque envelope. Dashed lines: constant-power hyperbolas.}
	\label{fig:engine_map}
\end{figure}

The Willans line expresses the fuel mass flow as
\begin{equation}
	\dot{m}_f = \dot{m}_{f,0}(n) + k_f(n)\, P
\end{equation}
\noindent where $\dot{m}_{f,0}(n) = c_1 + c_2\,n' + c_3\,n'^{2}$ captures speed-dependent friction losses and $k_f(n) = c_4 + c_5\,n'$ is the incremental fuel cost per unit power, with $n' = n/1000$. The five coefficients were determined by ordinary least-squares from the design matrix $[\,1,\; n',\; n'^2,\; P,\; n'\!\cdot\!P\,]$ applied to the ten crankshaft-corrected calibration points. The resulting engine map (\cref{fig:engine_map}) reproduces the measured SFC from \SI{208}{\gram\per\kilo\watt\per\hour} at \SI{1400}{\rpm} full load to \SI{272}{\gram\per\kilo\watt\per\hour} at \SI{1800}{\rpm} and \SI{40}{\percent} load, a \SI{64}{\gram\per\kilo\watt\per\hour} spread driven by both speed and load that the ASAE~D497.7 fuel model cannot capture.

\begin{figure}[h]
	\centering
	\includegraphics[width=\columnwidth]{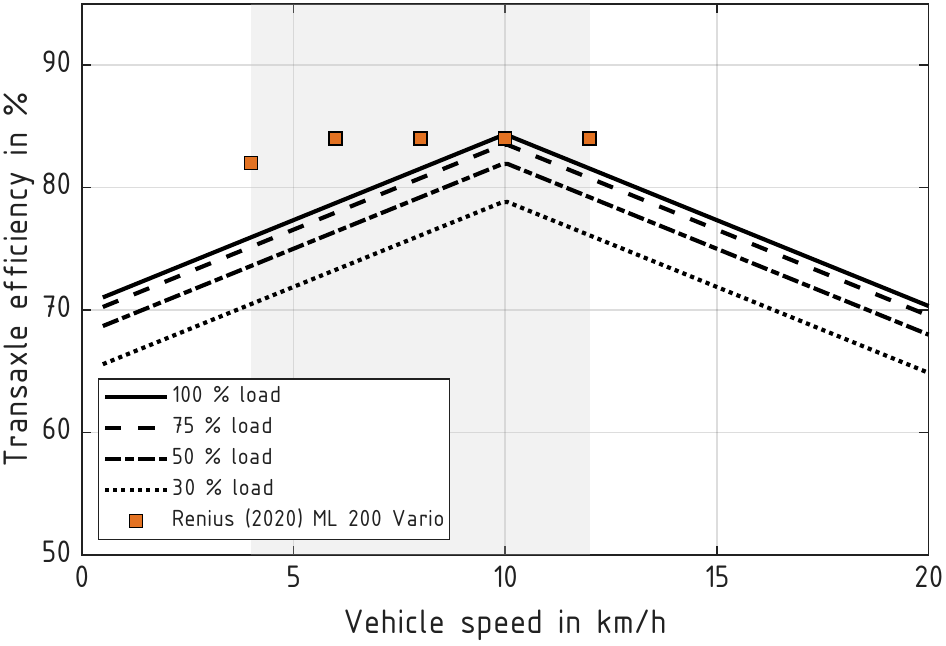}
	\caption{Transmission efficiency versus vehicle speed at four load levels. Square markers show the measured Fendt Vario ML~200 full-load data from Renius \cite{Renius2020} (fig.~5.69). The shaded region indicates the tillage working range.}
	\label{fig:trans_eff}
\end{figure}

\textbf{Transmission efficiency.} An analytical model based on the Kress \cite{Kress1968} power-split framework was implemented for an OC CVT structure. The hydrostatic power fraction is computed as $f_\mathrm{hydro} = |1 - v/v_\mathrm{sync}|$, where $v_\mathrm{sync} = \SI{10}{\kilo\metre\per\hour}$ is the synchronous (lock-up) speed of the primary range. This value was inferred from the measured efficiency peak in Renius \cite{Renius2020} (fig.~5.69) and is consistent with calculated losses for the ZF Terramatic~11 that show minimum losses at \SI{10}{\kilo\metre\per\hour} \cite{Ziegler2014}. Both major CVT architectures (OC and IC) place a lock-up point near the primary tillage speed range. Component efficiencies (spur gear $0.99$, planetary $0.98$, variator $0.85$, final drive $0.945$) were taken from Reick \cite{Reick2018}. The model was validated against the Fendt Vario ML~200 full-load data from Renius (fig.~5.69: \SI{82}{\percent} at \SI{4}{\kilo\metre\per\hour}, \SI{84}{\percent} at \SIrange{6}{12}{\kilo\metre\per\hour}) and against partial-load trends from the Reiter stepped-transmission map \cite{Reiter1990}. \Cref{fig:trans_eff} shows the resulting efficiency curves.

\begin{figure}[h]
	\centering
	\includegraphics[width=\columnwidth]{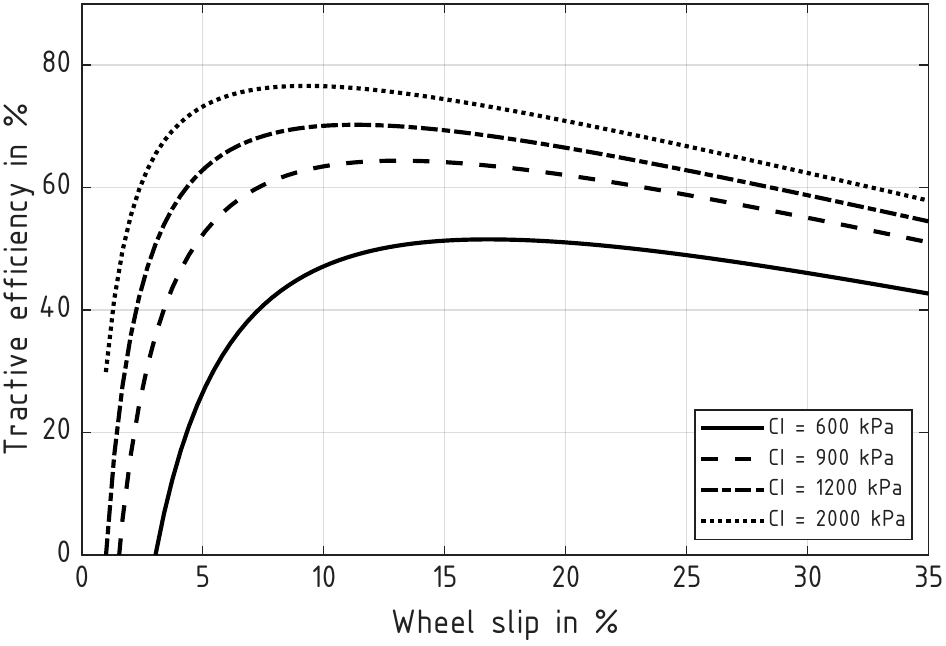}
	\caption{Tractive efficiency versus wheel slip for four soil cone-index levels. Radial tyre corrections from Zoz and Grisso \cite{ZozGrisso2003}.}
	\label{fig:traction}
\end{figure}

\textbf{Traction.} The Brixius model with Zoz and Grisso \cite{ZozGrisso2003} radial tyre corrections was implemented for a 4WD tractor with Michelin MultiBib tyres (front 540/65~R28, rear 650/65~R38) at a ballasted test weight of \SI{8540}{\kilo\gram} \cite{DLG2016PowerMix}. Multi-pass correction ($k_\mathrm{mp} = 1.1$, \cite{Renius2020}) accounts for the rear axle running on the pre-compacted track of the front axle. Slip is solved by bisection to match the required drawbar pull at each operating point. \Cref{fig:traction} shows tractive efficiency versus slip for four soil conditions.

\begin{figure}[h]
	\centering
	\includegraphics[width=\columnwidth]{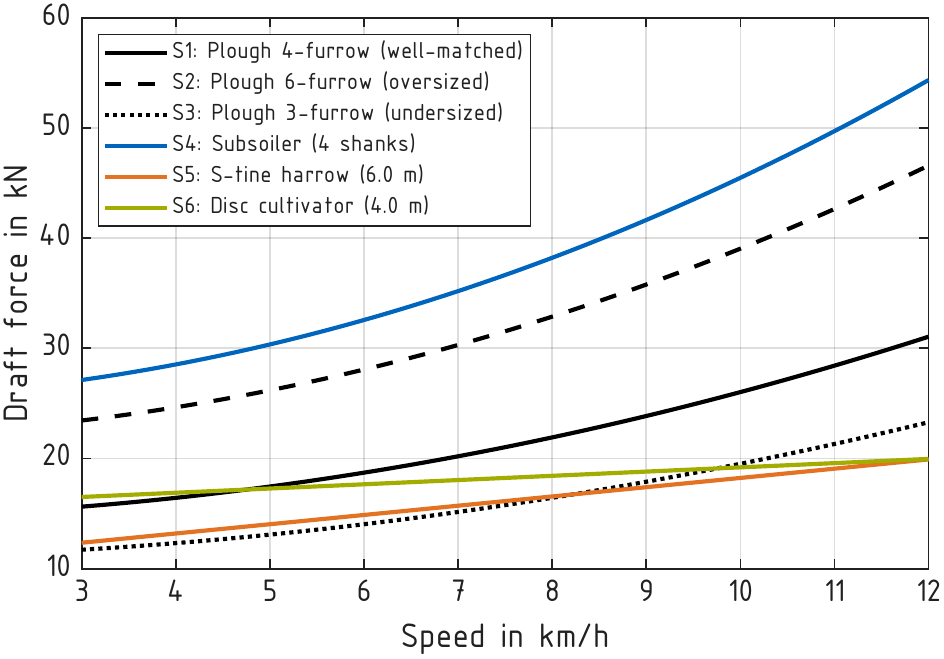}
	\caption{Draft force versus speed for all six tillage implements. The moldboard plough (S1 to S3) shows strong quadratic speed dependence ($C = 5.1$), while secondary tillage implements show linear speed dependence.}
	\label{fig:draft_force}
\end{figure}

\textbf{Draft force.} ASAE~D497.7 \cite{ASAED497} coefficients were used for
the moldboard plough ($A = 652$, $C = 5.1$)
and subsoiler ($A = 226$, $C = 1.8$).
For the European S-tine harrow, per-tine draft data from Ahokas and
Oksanen~\cite{Ahokas2015} (\SIrange{80}{182}{\newton\per tine} depending on
soil type at \SI{5}{\centi\metre} depth and \SI{8}{\kilo\metre\per\hour}) was
converted to equivalent area-based coefficients
($A = 228$, $B = 29$) that
include tine draft plus accessory resistance (front board, roller). The
speed-dependent term~$B$ reflects the significant increase in per-tine draft
with speed reported by Ahokas (table~2.1 in \cite{Ahokas2015}). For the
compact disc cultivator (European type, not American tandem disc), the draft
coefficient was back-calculated from manufacturer power requirements
(\SI{22}{\kilo\watt\per\metre} at \SIrange{10}{14}{\kilo\metre\per\hour}),
yielding $A = 480$ with $B = 12$.
\Cref{fig:draft_force} shows draft force versus speed for all six implements.

The submodels employ synthetic efficiency maps whose topology and gradient orientations are consistent with published empirical data. For the purposes of this proof of concept, the absolute accuracy of individual submodel values is of secondary importance; the essential requirement is that the cumulative efficiency along the drivetrain exhibits operating-point dependence, thereby providing a non-trivial optimization landscape that the solver can exploit. The fidelity of these models is sufficient to capture the relevant variability required to validate the EcoTIM concept.

\subsection{Test track and scenarios}
\label{subsec:track_scenarios}

\begin{figure}[h]
	\centering
	\includegraphics[width=\columnwidth]{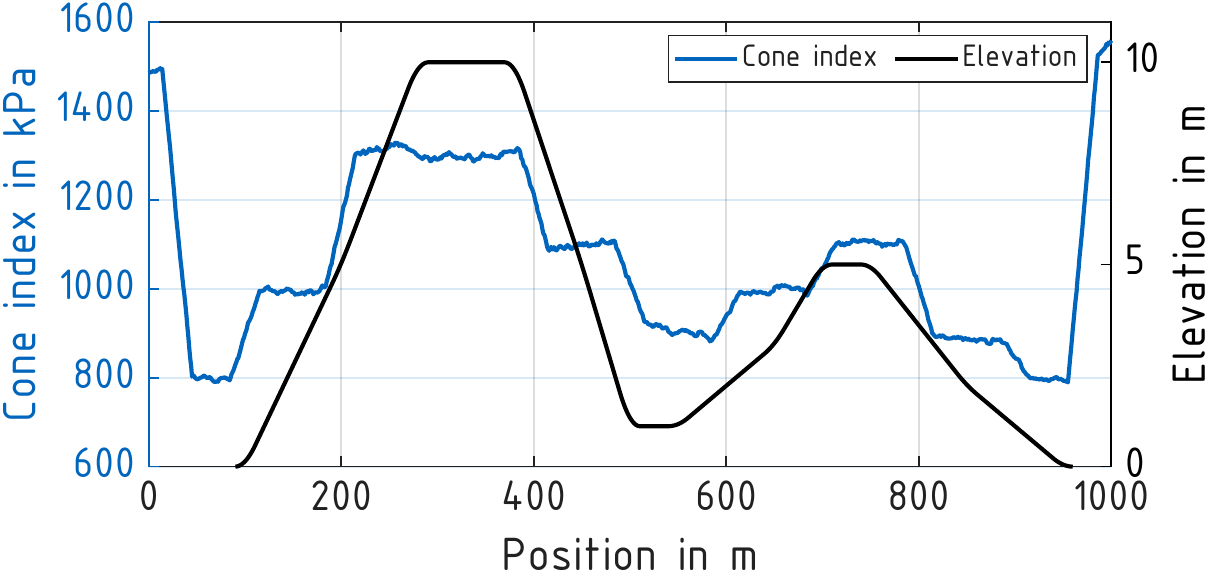}
	\caption{Virtual \SI{1}{\kilo\metre} test track: cone index (left axis) and elevation (right axis) versus position. Eight soil and terrain zones with smooth transitions.}
	\label{fig:track_profile}
\end{figure}

A \SI{1}{\kilo\metre} virtual test track was constructed with ten soil zones (\SI{100}{\metre} each) and a smooth elevation profile representative of a European arable field (\cref{fig:track_profile}). 
Five soil types (CI from \SIrange{800}{1300}{\kilo\pascal}) are each assigned to two zones, with firmer soils on hilltops and softer soils in the valley, reflecting the natural drainage pattern. An elevation profile with two hills provides grade variations of up to $\pm$\SI{8}{\percent} (approximately \SI{10}{\metre} total elevation range). The track is generated deterministically (fixed random seed) for reproducibility. Road driving and headland turns are out of scope.

Six scenarios were defined, each representing a \SI{100}{\kilo\watt} tractor with a different tillage implement sized for realistic European practice. \Cref{tab:scenarios} summarises the parameters. Scenarios~S1 to S3 test implement sizing effects (well-matched, oversized, undersized plough). Scenarios~S4 to S6 test different implement types (subsoiler, S-tine harrow, disc cultivator), all well-matched to the \SI{100}{\kilo\watt} tractor. The operator-defined speed limits ($v_\mathrm{min}$, $v_\mathrm{max}$) are used by the EcoTIM optimizer to clamp its speed commands. The minimum speed $v_\mathrm{min} = \SI{4}{\kilo\metre\per\hour}$ for most scenarios prevents the optimizer from reducing speed to near-zero in pursuit of fuel savings, which would harm productivity. For the subsoiler, $v_\mathrm{min} = \SI{2.5}{\kilo\metre\per\hour}$ allows slower operation because deep loosening is inherently a low-speed process.

\begin{table*}[t]
\centering
\caption{Simulation scenarios. All scenarios use the same \SI{100}{\kilo\watt} tractor and \SI{1}{\kilo\metre} test track. $v_\mathrm{BL}$ is the scenario-specific operator-selected baseline speed (Eco mode) used for the comparison in \cref{tab:fuel_comparison}; speeds are in \si{\kilo\metre\per\hour}.}
\label{tab:scenarios}
\small
\begin{tabular}{clcccccccc}
\hline
\# & Implement & $w_\mathrm{eff}$ [m] & $T$ [cm] & $v_\mathrm{min}$ & $v_\mathrm{BL}$ & $v_\mathrm{max}$ & $A$ & $B$ & $C$ \\
\hline
S1 & Plough 4-furrow (well-matched) & 1.6 & 20 & 4 &  8 & 10 & 652 & 0 & 5.1 \\
S2 & Plough 6-furrow (oversized)    & 2.4 & 20 & 4 &  8 & 10 & 652 & 0 & 5.1 \\
S3 & Plough 3-furrow (undersized)   & 1.2 & 20 & 4 & 10 & 10 & 652 & 0 & 5.1 \\
S4 & Subsoiler (4 shanks)           & 2.0 & 40 & 2.5 &  6 &  8 & 226 & 0 & 1.8 \\
S5 & S-tine harrow (6~m)            & 6.0 &  5 & 4 & 12 & 12 & 328 & 28 & 0 \\
S6 & Disc cultivator (4~m)          & 4.0 &  8 & 4 & 12 & 12 & 480 & 12 & 0 \\
\hline
\end{tabular}
\end{table*}

\subsection{Results}
\label{subsec:sim_result}

\textbf{Submodel characteristics.} \Cref{fig:engine_map} shows the engine SFC map (engine speed versus torque) with iso-consumption lines calibrated to the DLG data. The best SFC (\SI{223}{\gram\per\kilo\watt\per\hour}) occurs at the maximum-torque point (\SI{1400}{\rpm}, full load), which is characteristic of modern VGT common-rail turbodiesels where high boost at low speed pushes the efficiency peak onto the full-load boundary. \Cref{fig:trans_eff} shows transmission efficiency versus vehicle speed for four load levels, with the Fendt Vario validation points from Renius (fig.~5.69) overlaid. The model reproduces the measured \SIrange{82}{84}{\percent} efficiency across the \SIrange{4}{12}{\kilo\metre\per\hour} tillage range. \Cref{fig:traction} shows tractive efficiency versus slip for four cone-index levels, confirming the expected \SIrange{65}{70}{\percent} optimum at \SIrange{8}{12}{\percent} slip on tilled soil. \Cref{fig:draft_force} shows draft force versus speed for all six implements, illustrating the strong quadratic speed dependence of the plough ($C = 5.1$) versus the linear speed dependence of the S-tine harrow and disc cultivator.

\begin{figure*}[t]
	\centering
	\includegraphics[width=\textwidth]{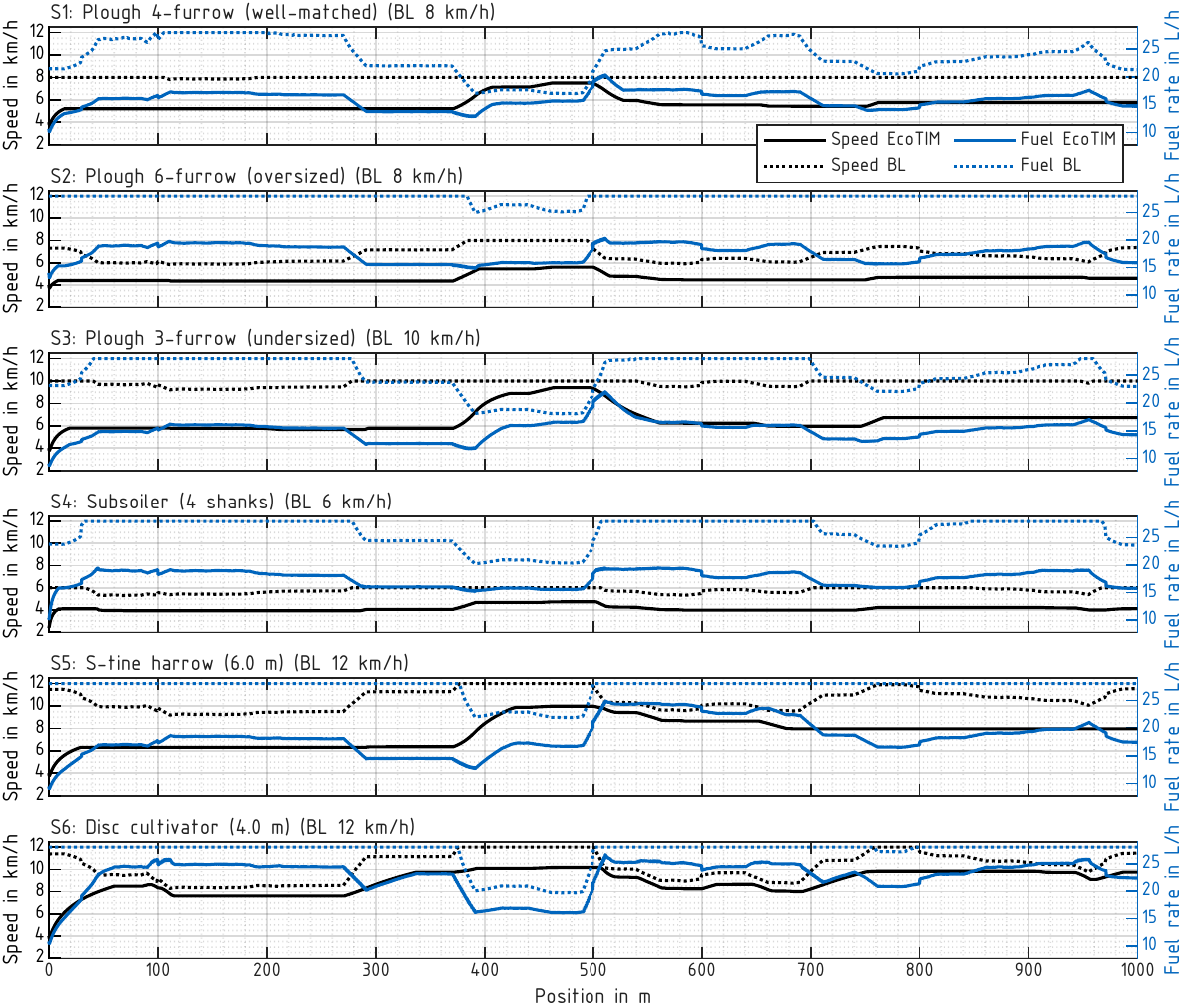}
	\caption{EcoTIM \SI{1}{\kilo\metre} field pass for all six scenarios. Each subplot shows vehicle speed (left axis, \si{\kilo\metre\per\hour}, black) and fuel rate (right axis, \si{\litre\per\hour}, blue) versus track position. Solid lines: EcoTIM enabled. Dotted lines: baseline at the scenario-specific operator speed (actual speed shown, which drops where the engine cannot sustain the setpoint). The baseline speed for each scenario is indicated in the subplot title. All subplots share identical vertical scales for direct comparison.}
	\label{fig:field_pass_overview}
\end{figure*}

\textbf{Field pass behaviour.} \Cref{fig:field_pass_overview} shows the \SI{1}{\kilo\metre} field pass for all six scenarios, comparing EcoTIM (solid lines) with the baseline at the scenario-specific operator speed (dotted lines). For each scenario, both vehicle speed (left axis, black) and instantaneous fuel rate (right axis, blue) are plotted against track position. The EcoTIM speed traces show that the optimizer adapts continuously to local conditions: reducing speed in the damp hollow (low CI, high draft, poor traction) and increasing speed on the firm stubble zone (high CI, better traction). For heavy implements (S1 to S4), the baseline tractor cannot always maintain the nominal speed in demanding zones and the actual speed drops, which is visible in the dotted speed curves. The EcoTIM optimizer anticipates these conditions and adjusts speed proactively rather than reactively.

For the baseline runs, the instantaneous fuel per hectare varies significantly along the track even at constant speed, because the changing soil conditions and grade affect engine load, traction efficiency, and (where the engine overloads) the actual achievable speed. This variation is captured by the position-resolved simulation.

\begin{figure}[h]
	\centering
	\includegraphics[width=\columnwidth]{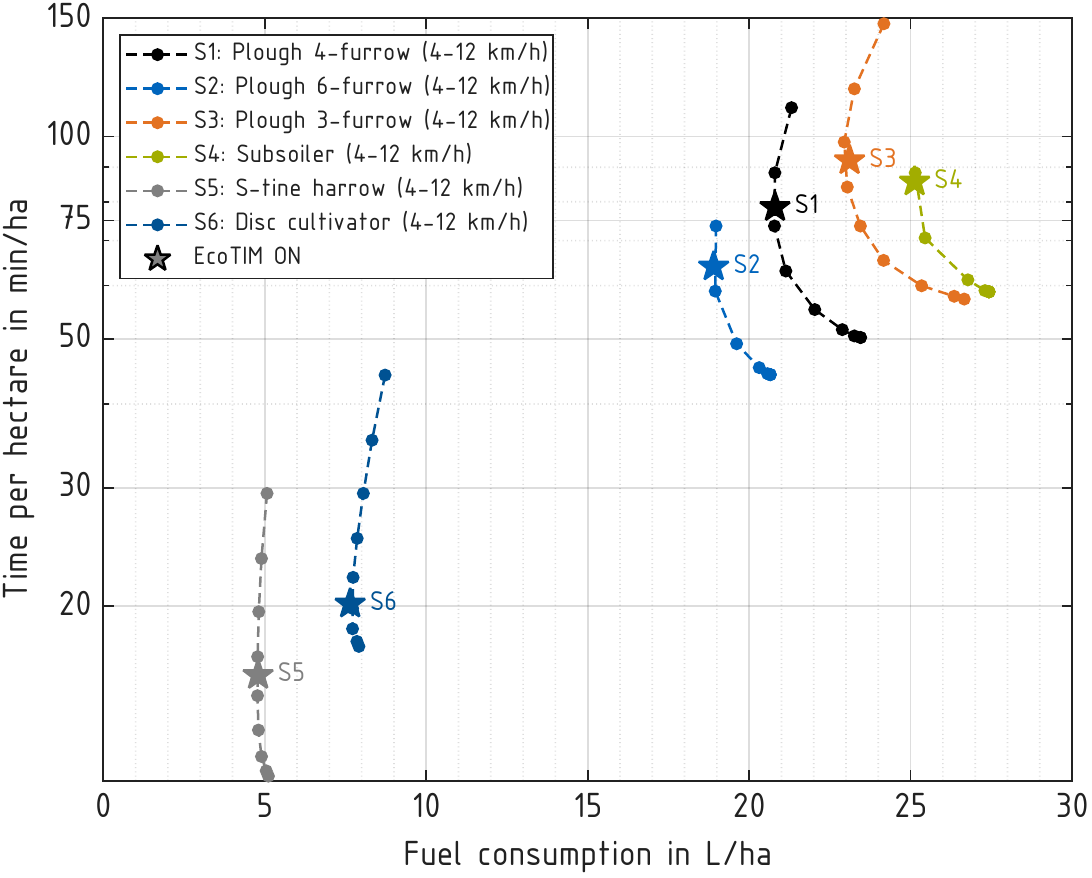}
	\caption{Fuel consumption (\si{\litre\per\hectare}, linear axis) versus time per hectare (\si{\minute\per\hectare}, logarithmic axis) for all six scenarios. Dashed lines with circles: baseline at nine constant speed setpoints (\SIrange{4}{12}{\kilo\metre\per\hour}). Stars: EcoTIM optimiser result. Each scenario is colour-coded and labelled. Heavy implements (S1 to S4) occupy the upper right; light, wide implements (S5, S6) occupy the lower left.}
	\label{fig:fuel_vs_time}
\end{figure}

\textbf{Fuel versus productivity.} \Cref{fig:fuel_vs_time} shows fuel consumption (\si{\litre\per\hectare}) versus work time (\si{\minute\per\hectare}) for all six scenarios on a semi-logarithmic scale (linear fuel, logarithmic time).

Each scenario appears as a dashed curve of nine baseline points (constant speed setpoints \SIrange{4}{12}{\kilo\metre\per\hour}) and a single EcoTIM star marking the optimizer result. The baseline curves trace the inherent fuel-productivity trade-off for each implement: heavy implements (plough, subsoiler) occupy the upper-right region (high fuel, slow), while light, wide implements (S-tine harrow, disc cultivator) occupy the lower-left region.

\Cref{tab:fuel_comparison} quantifies the comparison between EcoTIM and the scenario-specific baseline speed that a skilled operator would typically select (for example, \SI{8}{\kilo\metre\per\hour} for a well-matched 4-furrow plough, \SI{12}{\kilo\metre\per\hour} for a light disc cultivator). 

\begin{table*}[b]
	\centering
	\caption{Fuel consumption and time per hectare: scenario-specific baseline speed (operator-selected, Eco mode) versus EcoTIM enabled. Fuel savings are relative to the baseline. The time increase reflects the optimizer adapting speed to local soil conditions.}
	\label{tab:fuel_comparison}
	\small
	\begin{tabular}{clccccccc}
		\hline
		& & & \multicolumn{3}{c}{Fuel consumption in \si{\litre\per\hectare}} & \multicolumn{3}{c}{Time in \si{\minute\per\hectare}} \\
		\cmidrule(lr){4-6} \cmidrule(lr){7-9}
		\# & Implement & $v_\mathrm{BL}$ & Baseline & EcoTIM & $\Delta$ & Baseline & EcoTIM & $\Delta$ \\
		\hline
		S1 & Plough 4-furrow (well-matched) & 8  & 22.0 & 20.8 & $-$5.6\,\% & 55.3 & 78.5 & $+$42\,\% \\
		S2 & Plough 6-furrow (oversized)    & 8  & 20.6 & 18.9 & $-$8.1\,\% & 44.4 & 64.0 & $+$44\,\% \\
		S3 & Plough 3-furrow (undersized)   & 10 & 25.3 & 23.1 & $-$8.8\,\% & 60.0 & 92.2 & $+$54\,\% \\
		S4 & Subsoiler (4 shanks)           & 6  & 26.8 & 25.1 & $-$6.2\,\% & 61.2 & 85.7 & $+$40\,\% \\
		S5 & S-tine harrow (6.0~m)          & 12 &  5.1 &  4.8 & $-$6.3\,\% & 11.2 & 15.8 & $+$41\,\% \\
		S6 & Disc cultivator (4.0~m)        & 12 &  7.9 &  7.6 & $-$3.4\,\% & 17.4 & 20.2 & $+$16\,\% \\
		\hline
	\end{tabular}
\end{table*}

EcoTIM reduces fuel consumption in every scenario, with savings ranging from \SI{3.4}{\percent} (disc cultivator, S6) to \SI{8.8}{\percent} (undersized 3-furrow plough, S3). The largest savings occur where the baseline speed deviates most from the system optimum: the undersized 3-furrow plough at \SI{10}{\kilo\metre\per\hour} (S3, \SI{8.8}{\percent}) and the oversized 6-furrow plough at \SI{8}{\kilo\metre\per\hour} (S2, \SI{8.1}{\percent}), where the operator's habitual speed is far from the combined-efficiency minimum. Even for the well-matched 4-furrow plough (S1), where the baseline speed is already close to the optimum, EcoTIM still achieves a \SI{5.6}{\percent} fuel reduction by continuously adapting speed to the spatially varying soil conditions along the track.

These fuel savings come at the cost of increased time per hectare, ranging from $+$\SI{16}{\percent} (S6) to $+$\SI{54}{\percent} (S3), because the optimizer consistently favours lower speeds where the combined tractor-implement efficiency is highest. In \cref{fig:fuel_vs_time}, the EcoTIM stars lie close to the baseline Pareto curves, confirming that the optimizer exploits the same fundamental fuel-time trade-off but selects a more fuel-efficient operating region than the operator's default. Importantly, the fuel savings are achieved without any hardware modification and without requiring the operator to know the optimal speed for each soil zone; the implement ECU determines it automatically from the broadcast efficiency signal. In practice, the operator retains full control over the acceptable speed range ($v_\mathrm{min}$, $v_\mathrm{max}$) and can tighten the bounds to limit the productivity penalty when timeliness is critical.

\begin{figure*}[t]
	\centering
	\includegraphics[width=\textwidth]{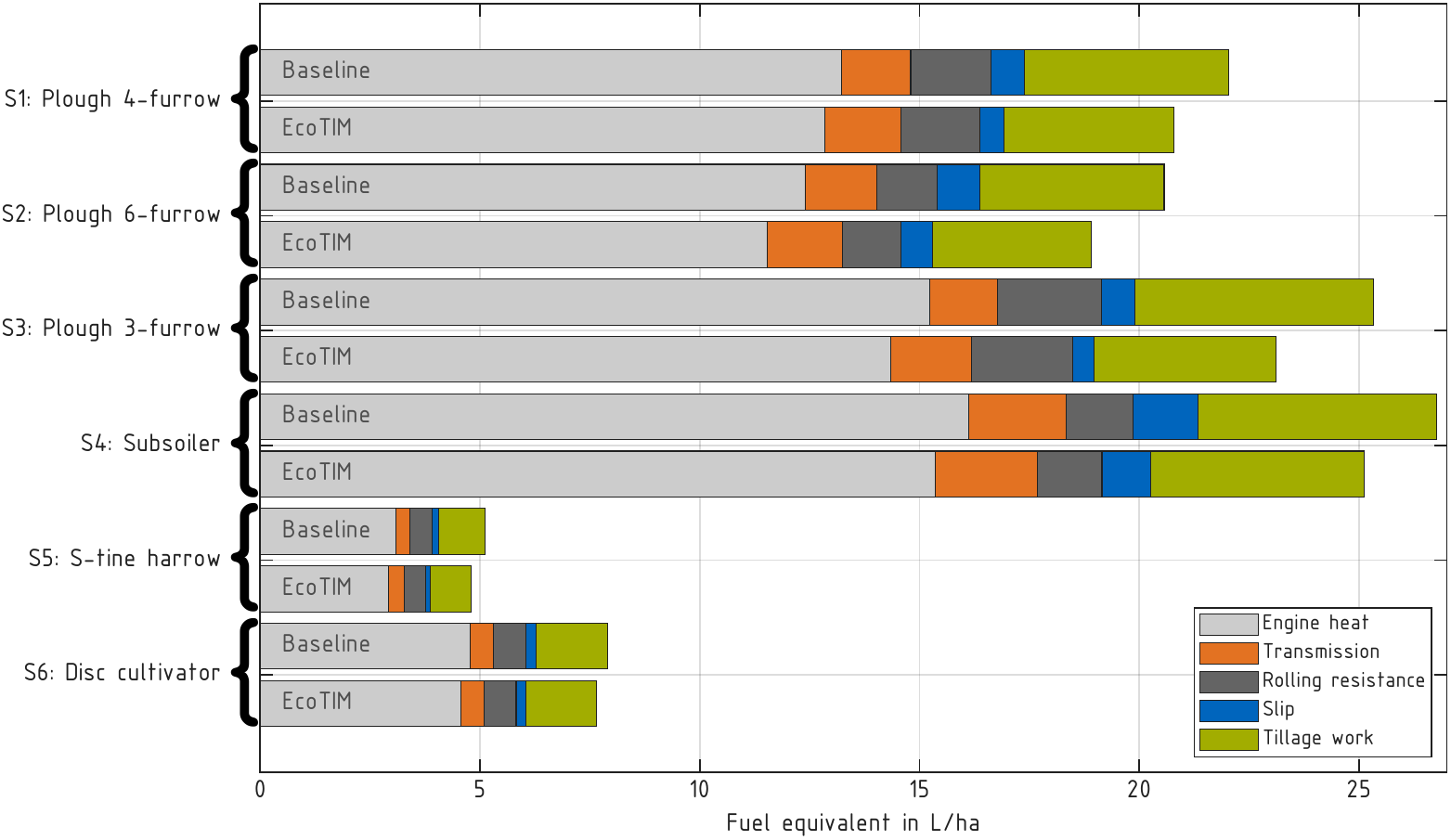}
	\caption{Energy balance per hectare for all six scenarios, comparing the scenario-specific baseline speed with EcoTIM enabled. Each horizontal bar shows the fuel energy partitioned into engine thermal losses, transmission losses, rolling resistance, wheel slip, and useful tillage work, expressed as fuel equivalent in \si{\litre\per\hectare}.}
	\label{fig:energy_balance}
\end{figure*}

\textbf{Energy balance.} \Cref{fig:energy_balance} shows the energy balance per hectare for all six scenarios, comparing the scenario-specific baseline speed with EcoTIM enabled. 

Each pair of horizontal bars shows how the fuel chemical energy is partitioned into engine thermal losses, transmission losses, rolling resistance, wheel slip, and useful tillage work, expressed as fuel equivalent in \si{\litre\per\hectare}. The energy balance closes to within rounding error for all scenarios, confirming internal consistency of the simulation.

For heavy implements (S1 to S4), the total fuel energy is dominated by engine thermal losses (approximately \SI{61}{\percent}) and the tillage-work fraction remains below \SI{20}{\percent}. The EcoTIM bars are consistently shorter than the corresponding baseline bars, confirming that the optimizer reduces total fuel input while delivering comparable tillage work. The relative size of the loss components shifts only slightly between baseline and EcoTIM, indicating that the savings arise primarily from operating the entire drivetrain at a more favourable load point rather than from improving any single subsystem. For the lighter implements (S5, S6), the absolute fuel consumption is much lower, but the energy partitioning follows the same pattern.

The subsystem efficiencies from the simulation fall within the ranges reported
in the literature. Engine thermal efficiency spans \SIrange{19}{40.5}{\percent}
(Willans line calibrated to 10 DLG PowerMix operating points
\cite{DLG2016PowerMix}, corrected from PTO to crankshaft). The upper bound
matches the best-point BSFC of \SI{\approx 205}{\gram\per\kilo\watt\per\hour}
($\eta \approx \SI{41}{\percent}$) read from the Deutz TCD~2013 L06~4V fuel map
(Fig.~4.23 in \cite{Renius2020}) and the \SI{\approx 208}{\gram\per\kilo\watt\per\hour}
reported for the Deutz TCD~4.1 L4 Agri \cite{Deutz_TCD41L4}; the lower bound
corresponds to low-torque part-load operation along the
\SIrange{280}{335}{\gram\per\kilo\watt\per\hour} contours encountered during
idle, headland turns, and light transport.

Transmission efficiency spans \SIrange{76}{85}{\percent}, consistent with the
\SIrange{82}{84}{\percent} reported at full load across
\SIrange{4}{12}{\kilo\metre\per\hour} for the Fendt Vario (Fig.~5.69 in
\cite{Renius2020}); the lower end reflects heavy-load operation below the
synchronous lock-up speed of \SI{10}{\kilo\metre\per\hour}, where the
hydrostatic power fraction increases. Tractive efficiency spans
\SIrange{60}{65}{\percent}, slightly below the \SIrange{65}{75}{\percent}
reported on firm tilled soil \cite{ZozGrisso2003}, owing to the damp hollow
zone ($\mathrm{CI} \approx \SI{800}{\kilo\pascal}$) in the test track.

The resulting fuel-to-tillage efficiency of \SIrange{17}{20}{\percent} follows
as the product of the three sub-efficiencies, i.e.\ approximately
\SI{80}{\percent} of the diesel chemical energy is dissipated as heat and
friction. Per-hectare fuel consumption was cross-validated against German
practice data \cite{KTBL2018} (rule of thumb: \SI{0.8}{\litre\per\hectare} per
cm of working depth), with all six scenarios falling within the expected ranges.

\section{Discussion}
\label{sec:discussion}

\subsection{Model validity and limitations}
\label{subsec:model_validity}

The submodels employed in this proof of concept use realistic but fictive efficiency maps whose shapes and gradient directions are consistent with published data from Renius \cite{Renius2020}, ASAE~D497.7 \cite{ASAED497}, and DLG PowerMix measurements. The absolute values of individual submodel outputs (for example, whether the transmission efficiency at \SI{7}{\kilo\metre\per\hour} is \SI{82}{\percent} or \SI{86}{\percent}) affect the computed litres per hectare but do not determine whether EcoTIM finds a better operating point. The optimizer relies on the derivative $\mathrm{d}\eta/\mathrm{d}v$, not on the absolute value of $\eta_\text{tractor}$. Therefore, the concept validity does not depend on exact calibration of the submodel parameters.

The engine SFC map, calibrated to 10 DLG operating points via the Willans line, places the minimum $b_e$ at the maximum-torque point (\SI{1400}{\rpm}, full load). The PTO-to-crankshaft correction ($\eta_\mathrm{PTO} = 0.93$) is an aggregate factor anchored to two published crankshaft references (Renius \cite{Renius2020}, fig.~4.23; Deutz TCD~4.1~L4~Agri \cite{Deutz_TCD41L4}). The PTO clutch is not modelled as a separate component; a tractor-specific measurement of the PTO drivetrain would refine this value. For the comparative analysis between baseline and EcoTIM, the relative differences between operating points matter more than the absolute island shape.

The transmission model uses a fictive synchronous speed of \SI{10}{\kilo\metre\per\hour} for the primary range, which was inferred from measured data for both OC \cite{Renius2020} and IC \cite{Ziegler2014} architectures. Proprietary CVT efficiency maps are not published by tractor manufacturers, which is a known gap in the literature. Nevertheless, the analytical model reproduces the measured full-load efficiency trend (\SIrange{82}{84}{\percent} across \SIrange{4}{12}{\kilo\metre\per\hour}) and the partial-load behaviour reported by Reiter \cite{Reiter1990}. The gradient direction (efficiency increasing toward the synchronous speed, decreasing away from it) is physically correct for any power-split CVT, regardless of the exact parameter values.

Tillage implement draft force models were taken from ASAE~D497.7 where available and supplemented with European data from Ahokas and Oksanen \cite{Ahokas2015} for the S-tine harrow and from manufacturer specifications for the compact disc cultivator. The ASAE model has a stated uncertainty of $\pm$\SI{25}{\percent} to $\pm$\SI{50}{\percent} depending on implement type \cite{ASAED497}. In practice, many tillage implements include additional elements beyond the primary soil-engaging tools, such as front boards, packer rollers, levelling bars, and depth wheels, which contribute to draft force but are not individually parameterised in the ASAE tabulation. In the S-tine harrow scenario, these accessories were included as an equivalent area-based coefficient. However, for more complex combined implements (for example, a cultivator with disc section, tine section, and packer in series), the simple ASAE model may not capture the total draft accurately, and implement-specific calibration would be required.

Tillage quality was not sensed or modelled. The operator sets the minimum and maximum speed limits to ensure acceptable work quality, and EcoTIM respects these limits. In practice, the relationship between travel speed and tillage quality depends on the implement type, soil conditions, and the specific quality criterion (e.g.\ aggregate size, surface roughness, residue incorporation). Quality sensing is considered future work.

The traction efficiency model requires the soil cone index~CI as an input. In the simulation, CI is provided directly from the scenario definition or the track profile. On a real tractor, CI must be estimated, for example from real-time slip-torque observation, on-board soil sensors, or geo-referenced prescription maps. The accuracy and update rate of CI estimation will affect the quality of the broadcast efficiency and its derivative. However, since the optimizer depends on the gradient direction $\mathrm{d}\eta/\mathrm{d}v$ rather than on the absolute value of $\eta_\text{tractor}$, moderate CI estimation error is expected to shift the optimum only slightly rather than invalidate the concept.

\subsection{Optimizer cycle time}
\label{subsec:optimizer_rate}

The proof-of-concept simulation runs the implement-side optimizer at every CAN bus tick (\SI{100}{\milli\second}). The optimizer algorithm (\cref{subsec:optimizer}) requires three draft-force evaluations and one linear interpolation per cycle, which is computationally trivial on any modern embedded processor. However, in a real implementation, sensor noise on the received $\eta_\text{tractor}$ and $\mathrm{d}\eta/\mathrm{d}v$ values, CAN bus latency, and tractor response dynamics (time constant $\tau \approx \SI{2}{\second}$) may require signal filtering that limits the effective update rate.

Since the tractor speed response time constant is approximately \SI{2}{\second}, the optimizer cannot meaningfully change speed faster than once per second, regardless of its update rate. Therefore, an optimizer cycle time of \SI{500}{\milli\second} to \SI{1}{\second} is likely sufficient for product implementation, while \SI{100}{\milli\second} provides smoother acceleration commands and better responsiveness to abrupt zone transitions. A systematic sensitivity study on optimizer rate versus convergence behaviour is recommended for product development. The ISOBUS TIM speed command update rate is independent, regardless of the optimizer rate, with the most recent speed setpoint held between updates.

\subsection{Backward compatibility and non-CVT transmissions}
\label{subsec:backwards_compat}

The EcoTIM message architecture was designed for backward compatibility. A non-EcoTIM implement connected to an EcoTIM tractor simply does not receive the efficiency broadcast, and the tractor operates normally. Conversely, a non-EcoTIM tractor connected to an EcoTIM implement does not transmit the efficiency broadcast; the implement detects this during the TIM session handshake and disables the optimizer. The acceleration extension to the TIM speed command uses reserved bytes that are ignored by non-EcoTIM tractors.

The simulation was conducted with a power-split CVT, which is the common transmission type in European tractors above \SI{100}{\kilo\watt}. However, the EcoTIM concept is not limited to CVT transmissions. Full-powershift transmissions can also benefit, since the tractor-side efficiency computation ($\eta_\mathrm{engine} \cdot \eta_\mathrm{trans} \cdot \eta_\mathrm{tractive}$) and the broadcast of $\eta_\text{tractor}$ and $\mathrm{d}\eta/\mathrm{d}v$ are independent of the transmission type. With a powershift transmission, the tractor's internal Eco mode selects the gear that minimises SFC for the current speed and load, and the transmission efficiency varies discretely with gear selection rather than continuously with CVT ratio. The implement-side optimizer does not need to know whether the tractor has a CVT or powershift; it uses only the broadcast efficiency values, which already incorporate the transmission type. Nevertheless, experimental validation with powershift transmissions is recommended as future work, since the discrete speed steps at gear shifting may affect optimizer stability. However, this is the common TIM speed command engineering challenge for TIM tractors even without EcoTIM. 

\subsection{Control loop stability and edge cases}
\label{subsec:stability}

The EcoTIM concept defines a sophisticated approach to cooperative fuel optimisation, but it is recognised that various edge cases may arise in practice where the simplified models do not fully capture real-world behaviour. The gradient-descent optimizer assumes that the fuel proxy $D/\eta$ is locally smooth and unimodal, which holds for the standard ASAE draft models tested in this study. However, for implements with non-standard draft characteristics (for example, a combined machine with a disc section, chisel section, and packer in series, each with different speed dependencies), the combined fuel proxy may exhibit multiple local minima or discontinuities that could cause the optimizer to oscillate or converge to a suboptimal point.

The tractor dynamics model uses a simple first-order response with feed-forward, which captures the dominant behaviour of a CVT under draft load. In practice, the tractor's speed response depends on engine turbo lag, CVT hydraulic transients, tyre deformation dynamics, and the soil response to changing slip. These higher-order effects may cause the actual speed to deviate from the modelled trajectory, particularly during rapid zone transitions. The acceleration command partially mitigates this by providing the tractor's powertrain controller with advance information about the intended speed change, but careful engineering of the control loop gains and filter parameters is necessary for generic plug-and-play operation across different tractor and implement combinations. Oksanen~\cite{Oksanen2010} proposed a human-supervised identification procedure at each tractor-implement connection to handle parameter variation across tractors in ISO~11783 closed-loop control. Although targeted at the harder auxiliary-valve/cylinder case, the same principle, startup calibration before closing the loop, could be applied to the speed response time constant $\tau$ to improve EcoTIM robustness across tractor brands.

Therefore, while the simulation demonstrates the feasibility and potential of the EcoTIM concept, the path from proof of concept to a robust industrial product requires systematic testing of edge cases, stability margins, and failure modes across a representative range of tractor-implement combinations and field conditions.

\subsection{Future work}
\label{subsec:future_work}

Several directions are identified for future development:

\textbf{Full-scale prototype and field validation.} The simulation results should be validated with a physical tractor-implement system equipped with EcoTIM-capable ECUs. Field tests on real agricultural land with varying soil types, moisture levels, and topography would quantify the actual fuel savings under conditions that include sensor noise, CAN bus latency, and real powertrain dynamics. 

\textbf{Quality sensing.} In this study, the operator sets the minimum speed as a proxy for tillage quality. A natural extension is to equip the implement with a quality sensor (for example, a soil-roughness camera \cite{RieglerNurscher2020} or an aggregate-size measurement system) that provides a real-time quality feedback signal. The optimizer could then jointly minimise fuel per hectare subject to a minimum quality constraint, replacing the static speed limit with an adaptive, condition-responsive quality bound.

\textbf{PTO-driven implements.} The current scope is limited to draft-only tillage implements. Implements that require PTO power (rotary harrows, rotary tillers, power harrows) have a fundamentally different energy balance in which the majority of energy enters through the PTO shaft rather than through the drawbar. PTO-driven implements require a fixed PTO speed set by the implement (typically \SI{540}{\rpm} or \SI{1000}{\rpm}), so the engine speed is effectively locked and the EcoTIM optimiser can modulate only the torque axis. The tractor-side broadcast of $\eta_\mathrm{tractor}$ and $\mathrm{d}\eta/\mathrm{d}v$ would be restricted to the subset of operating points that satisfy the PTO speed constraint, and the implement-side optimizer would command torque (via implement depth or engagement) rather than ground speed. Extending EcoTIM to PTO-driven implements would require additional messages for PTO efficiency and a modified optimizer that considers both drawbar and PTO power paths.

\textbf{Powershift and hybrid-electric transmissions.} The framework extends naturally to drivetrain architectures beyond hydraulic power-split CVTs. Powershift transmissions produce step-wise discontinuities in $\mathrm{d}\eta/\mathrm{d}v$ at gear boundaries, while hybrid-electric drivetrains introduce dependence of $\eta_\mathrm{tractor}(v)$ on the instantaneous mechanical-electrical power split. In both cases the broadcast interface remains unchanged: the additional complexity is absorbed by the tractor-side efficiency estimator rather than communicated to the implement.

\textbf{Multi-implement configurations.} European farming practice increasingly uses front-mounted implements (seed drills, fertiliser applicators) simultaneously with rear-mounted tillage tools. Extending EcoTIM to multi-implement configurations would require coordinating the speed preferences of two or more implements, which introduces a multi-objective optimisation problem.

\textbf{Efficiency abstraction.} The current EcoTIM concept requires the tractor to broadcast its absolute combined efficiency $\eta_\mathrm{tractor}$ as a percentage. Tractor manufacturers may be reluctant to expose a value that directly reveals their powertrain's true efficiency to competitors. It should be investigated whether the broadcast can be abstracted, for example by transmitting a normalised or scaled efficiency value rather than the absolute figure, without sacrificing the optimizer's ability to determine the correct speed gradient. Since the implement-side optimizer uses primarily the derivative $\mathrm{d}\eta/\mathrm{d}v$ and the ratio $F_\text{draft}/\eta_\text{tractor}$ rather than $\eta_\text{tractor}$ alone, there may be formulations that preserve the gradient information while concealing the absolute efficiency level. This is left for future work.

\section{Conclusion}
\label{sec:conclusion}

In this paper, the original novel EcoTIM concept was presented with engineering details. 

A tractor-implement system can optimise its combined fuel efficiency in real time through the exchange of only two scalar values: the tractor's fused efficiency $\eta_\mathrm{tractor}$ and its speed derivative $\mathrm{d}\eta/\mathrm{d}v$. No proprietary subsystem models need to be disclosed between tractor and implement manufacturers.

The required data exchange can be realised with one new message and one byte-level extension to the existing TIM speed command. We propose that these messages are standardised within ISO~11783 to enable industry-wide, non-proprietary adoption.

The acceleration setpoint, transmitted alongside the speed command, enables feed-forward powertrain control on the tractor side, which reduces response lag and improves transient smoothness compared with speed-only TIM commands.

The proof-of-concept simulation across six tillage scenarios and a \SI{1}{\kilo\metre} spatially varying test track confirms that the implement-side gradient-descent optimizer converges to operating points with lower fuel per hectare than the constant-speed baseline, while remaining within operator-defined speed limits. 

EcoTIM is architecturally multi-brand: the tractor broadcasts its efficiency state without knowing the implement type, and the implement commands speed without knowing the engine map or transmission type. This plug-and-play property is not achievable with centralised optimisation approaches.

\begin{ack}
\label{sec:acknowledgements}

The original idea of the EcoTIM concept was born in discussions between Prof. Timo Oksanen and VDMA Landtechnik in December 2020. It was formalized into a presentation in January–February 2021 and first presented to VDMA Landtechnik member companies in February 2021, including CLAAS, CNHi, Amazone, John Deere, Horsch, and Lemken, among others. From March 2021 to July 2022, a series of round-table discussions examined the concept's limitations and scope.
\par
The authors want to thank the VDMA Landtechnik staff (Dr. Buitkamp and Dr. Hipp), as well as the industry experts from the aforementioned companies, for their valuable feedback. Special thanks are also due to Prof. Karl Th. Renius for his insights into tractor internal efficiency principles and his foundational contributions to the implement-commands-tractor concept.
\end{ack}

\newpage
\bibliographystyle{IEEEtran}
\bibliography{literature/literature_ecotim}

@IEEEtranBSTCTL{IEEEexample:BSTcontrol,
  CTLuse_forced_etal       = {no},
  CTLmax_names_forced_etal = {6},
  CTLnames_show_etal       = {3},
  CTLuse_url               = {yes},
  CTLuse_doi               = {yes}
}

@misc{J1939,
  author       = {{SAE International}},
  title        = {{SAE} {J1939} -- Serial Control and Communications Heavy Duty Vehicle Network},
  year         = {2023},
  organization = {SAE International},
  address      = {Warrendale, PA, USA}
}

@misc{J1939-71,
  author       = {{SAE International}},
  title        = {{SAE} {J1939-71} -- Vehicle Application Layer},
  year         = {2023},
  organization = {SAE International},
  address      = {Warrendale, PA, USA}
}

@misc{ISO11783-1,
  author       = {{International Organization for Standardization}},
  title        = {{ISO} 11783-1:2017 -- Tractors and machinery for agriculture and forestry -- Serial control and communications data network -- Part 1: General standard for mobile data communication},
  year         = {2017},
  organization = {ISO},
  address      = {Geneva, Switzerland}
}

@misc{ISO11783-7,
  author       = {{International Organization for Standardization}},
  title        = {{ISO} 11783-7:2015 -- Tractors and machinery for agriculture and forestry -- Serial control and communications data network -- Part 7: Implement messages application layer},
  year         = {2015},
  organization = {ISO},
  address      = {Geneva, Switzerland}
}

@misc{ISO11783-9,
  author       = {{International Organization for Standardization}},
  title        = {{ISO} 11783-9:2012 -- Tractors and machinery for agriculture and forestry -- Serial control and communications data network -- Part 9: Tractor {ECU}},
  year         = {2012},
  organization = {ISO},
  address      = {Geneva, Switzerland}
}

@inproceedings{Oksanen2021,
  author    = {Oksanen, Timo and Auernhammer, Hermann},
  title     = {{ISOBUS} -- The Open High-Speed Network That Enables Smart Implements: How It Was, Is, and Will Be},
  booktitle = {ASABE Distinguished Lecture Series No. 43},
  year      = {2021},
  organization = {American Society of Agricultural and Biological Engineers},
  address   = {St. Joseph, MI, USA}
}

@inproceedings{Oksanen2010,
  author    = {Oksanen, Timo},
  title     = {Closed loop control over {ISO} 11783 network -- challenges of plug-and-play},
  booktitle = {IFAC Proceedings Volumes (IFAC-PapersOnline)},
  volume    = {43},
  number    = {26},
  year      = {2010},
  pages     = {169--174},
  doi       = {10.3182/20101206-3-JP-3009.00030}
}

@book{Renius2020,
  author    = {Renius, Karl Theodor},
  title     = {Fundamentals of Tractor Design},
  publisher = {Springer},
  address   = {Cham, Switzerland},
  year      = {2020},
  doi       = {10.1007/978-3-030-32804-7}
}

@book{Ahokas2015,
  author    = {Ahokas, Jukka and Oksanen, Timo},
  title     = {Maamekaniikka},
  year      = {2015},
  edition   = {2},
  publisher = {Helsingin yliopisto, Maataloustieteiden laitos},
  series    = {Julkaisuja / Maataloustieteiden laitos},
  number    = {40},
  isbn      = {978-951-51-0142-6}
}

@book{Guzzella2010,
	author    = {Lino Guzzella and Christopher H. Onder},
	title     = {Introduction to Modeling and Control of Internal Combustion Engine Systems},
	publisher = {Springer Berlin Heidelberg},
	year      = {2010},
	edition   = {2},
	isbn      = {978-3-642-10775-7},
	doi       = {10.1007/978-3-642-10775-7},
	pages     = {XII, 354}
}

@misc{ASAED497,
  author       = {{American Society of Agricultural and Biological Engineers}},
  title        = {{ASAE} {D497.7} {MAR2011} -- Agricultural Machinery Management Data},
  year         = {2011},
  organization = {ASABE},
  address      = {St.\ Joseph, MI, USA}
}

@incollection{ZozGrisso2003,
  author    = {Zoz, Frank M. and Grisso, Robert D.},
  title     = {Traction and Tractor Performance},
  booktitle = {ASAE Distinguished Lecture Series, No.\ 27},
  year      = {2003},
  publisher = {ASAE},
  address   = {St.\ Joseph, MI, USA}
}

@inproceedings{Brixius1987,
  author    = {Brixius, W. W.},
  title     = {Traction Prediction Equations for Bias Ply Tires},
  booktitle = {ASAE Paper No.\ 87-1622},
  year      = {1987},
  organization = {ASAE},
  address   = {St.\ Joseph, MI, USA}
}

@techreport{DLG2016PowerMix,
	author       = {{DLG e.V.}},
	title        = {{Fendt 514 Vario S4: DLG PowerMix Datasheet}},
	institution  = {DLG Test Center},
	address      = {Groß-Umstadt, Germany},
	year         = {2016},
	month        = oct,
	number       = {2016-00697},
	url          = {https://www.dlg-test.de},
	note         = {Test performed on AGCO Fendt GmbH tractor; includes transport and field work power, fuel, and urea consumption measurements. Accessed April 17, 2026.}
}

@misc{Deutz_TCD41L4,
	author       = {{DEUTZ AG}},
	title        = {{TCD 4.1 L4 (Agri)} -- Product Data Sheet},
	year         = {2024},
	howpublished = {\url{https://www.deutz.com/de/produkte/engine-finder/engine-detail/tcd-41-l4-agri/}},
	note         = {4-cylinder 4.0~L Stage~IV/V diesel engine, rated 126~kW at 1900~rpm, minimum $b_e = 208$~g/kWh. Engine family of the Fendt~514 Vario~S4 (Deutz 4038~cm$^3$). Accessed April 16, 2026.}
}

@inproceedings{Kress1968,
  author    = {Kress, James H.},
  title     = {Hydrostatic Power-Splitting Transmissions for Wheeled Vehicles: Classification and Theory of Operation},
  booktitle = {SAE Technical Paper 680549},
  year      = {1968},
  organization = {SAE International},
  doi       = {10.4271/680549}
}

@phdthesis{Reick2018,
	author       = {Reick, Benedikt},
	year         = {2018},
	title        = {{Methode zur Analyse und Bewertung von stufenlosen Traktorgetrieben mit mehreren Schnittstellen}},
	doi          = {10.5445/KSP/1000084168},
	publisher    = {{KIT Scientific Publishing}},
	isbn         = {978-3-7315-0815-1},
	issn         = {1869-6058},
	series       = {Karlsruher Schriftenreihe Fahrzeugsystemtechnik / Institut für Fahrzeugsystemtechnik},
	pagetotal    = {221},
	school       = {Karlsruher Institut für Technologie (KIT)},
	language     = {german},
	volume       = {64}
}

@phdthesis{Reiter1990,
  author  = {Reiter, Heribert},
  title   = {Verluste und {W}irkungsgrade bei {T}raktorgetrieben},
  school  = {Technische Universit{\"a}t M{\"u}nchen},
  year    = {1990},
  address = {Munich, Germany},
  series  = {Fortschritt-Berichte VDI, Reihe 14, Nr.\ 46},
  publisher = {VDI-Verlag},
  isbn    = {3-18-144614-9}
}

@inproceedings{Ziegler2014,
  author    = {Ziegler, J. and Bailly, G. and Pohlenz, J.},
  title     = {Rechnerische {V}erlustleistungsanalyse von stufenlosen {G}etriebesystemen am {B}eispiel des {ZF} {T}erramatic11},
  booktitle = {VDI-Berichte 2226},
  pages     = {423--431},
  year      = {2014},
  publisher = {VDI-Verlag},
  address   = {D{\"u}sseldorf, Germany}
}

@phdthesis{Pichlmaier2012,
  author  = {Pichlmaier, Benno},
  title   = {Traktionsmanagement f{\"u}r {T}raktoren},
  school  = {Technische Universit{\"a}t M{\"u}nchen},
  year    = {2012},
  address = {Munich, Germany},
  url     = {https://mediatum.ub.tum.de/1098891}
}

@article{Troesken2020,
  author  = {Tr{\"o}sken, Lennart and Meiners, Arwid and Frerichs, Ludger and B{\"o}ttinger, Stefan},
  title   = {Modellbasierte {B}erechnung von {K}raftstoffverbr{\"a}uchen landwirtschaftlicher {V}erfahrensketten},
  journal = {Landtechnik -- Agricultural Engineering},
  volume  = {75},
  number  = {4},
  year    = {2020},
  pages   = {278--299},
  doi     = {10.15150/lt.2020.3253}
}

@inproceedings{EKoTech2017,
  author    = {Decker, Max and Frerichs, Ludger and B{\"o}ttinger, Stefan and Remmele, Edgar},
  title     = {{EKoTech} -- {A} Holistic Approach to Reduce {CO\textsubscript{2}} Emissions of Agricultural Machinery in Process Chains},
  booktitle = {SAE Technical Paper 2017-01-1929},
  year      = {2017},
  organization = {SAE International},
  doi       = {10.4271/2017-01-1929}
}

@article{Roessler2012,
	author    = {R\"{o}\ss{}ler, Patrick and Kautzmann, Timo and Geimer, Marcus},
	title     = {Online configurable tractor implement models},
	journal   = {Landtechnik -- Agricultural Engineering},
	year      = {2012},
	volume    = {67},
	number    = {4},
	pages     = {247--250},
	doi       = {10.15150/lt.2012.303}
}

@book{KTBL2018,
	author       = {{Kuratorium für Technik und Bauwesen in der Landwirtschaft (KTBL)}},
	title        = {Faustzahlen für die {L}andwirtschaft},
	edition      = {15},
	publisher    = {Kuratorium für Technik und Bauwesen in der Landwirtschaft},
	address      = {Darmstadt},
	year         = {2018},
	isbn         = {978-3-941583-64-2},
	language     = {german},
}

@article{RieglerNurscher2020,
	title = {Machine vision for soil roughness measurement and control of tillage machines during seedbed preparation},
	journal = {Soil and Tillage Research},
	volume = {196},
	pages = {104351},
	year = {2020},
	issn = {0167-1987},
	doi = {https://doi.org/10.1016/j.still.2019.104351},
	author = {Peter Riegler-Nurscher and Gerhard Moitzi and Johann Prankl and Josef Huber and Jürgen Karner and Helmut Wagentristl and Markus Vincze}
}

@ARTICLE{Soitinaho2023,
	author={Soitinaho, Riikka and Oksanen, Timo},
	journal={IEEE Transactions on Control Systems Technology}, 
	title={Local Navigation and Obstacle Avoidance for an Agricultural Tractor With Nonlinear Model Predictive Control}, 
	year={2023},
	volume={31},
	number={5},
	pages={2043-2054},
	doi={10.1109/TCST.2023.3291533}}

\end{document}